\documentclass[draftclsnofoot]{IEEEtran}
\usepackage{graphicx}
\usepackage{epstopdf}

\usepackage{subfigure}
\usepackage{booktabs}
\usepackage{caption}
\usepackage{color}
\usepackage{bm}
\usepackage{CJK}
\usepackage{indentfirst}
\usepackage{amsmath}
\usepackage{amssymb}
\usepackage{float}
\usepackage{multirow}
\usepackage{algorithm}
\usepackage{algpseudocode}
\usepackage{amsmath}
\usepackage{flushend}
\usepackage{geometry}
\newtheorem{assumption}{Assumption}

\begin{document}
\title{Channel Customization for Limited Feedback in RIS-assisted FDD Systems}
\author{\IEEEauthorblockN{Weicong~Chen, \emph{Graduate Student Member, IEEE}, Chao-Kai~Wen, \emph{Senior Member IEEE,} Xiao~Li, \emph{Member, IEEE,} Michail~Matthaiou, \emph{Fellow, IEEE} and~Shi~Jin, \emph{Senior Member, IEEE}}\\
\thanks{
{W. Chen, X. Li, and S. Jin are with the National Mobile Communications
Research Laboratory, Southeast University, Nanjing 210096, China (e-mail: cwc@seu.edu.cn; li\_xiao@seu.edu.cn; jinshi@seu.edu.cn).} }
\thanks{C.-K. Wen is with the Institute of Communications Engineering, National Sun Yat-sen University, Kaohsiung 80424, Taiwan. (e-mail: chaokai.wen@mail.nsysu.edu.tw). }
\thanks{M. Matthaiou is with the Centre for Wireless Innovation (CWI), Queen's University Belfast, Belfast BT3 9DT, U.K. (e-mail: m.matthaiou@qub.ac.uk).}
}

\maketitle
\begin{abstract}
Reconfigurable intelligent surfaces (RISs) represent a pioneering technology to realize smart electromagnetic environments by reshaping the wireless channel. \textcolor[rgb]{0,0,0}{Jointly designing the transceiver and RIS relies on the channel state information (CSI), whose feedback has not been investigated in multi-RIS-assisted frequency division duplexing systems.} In this study, the limited feedback of the RIS-assisted wireless channel is examined by capitalizing on the ability of the RIS in channel customization. \textcolor[rgb]{0,0,0}{By configuring the phase shifters of the surfaces using statistical CSI, we customize a sparse channel in rich-scattering environments, which significantly reduces the feedback overhead in designing the transceiver and RISs. Since the channel is customized in terms of singular value decomposition (SVD) with full-rank, the optimal SVD transceiver can be approached without a matrix decomposition and feeding back the complete channel parameters. The theoretical spectral efficiency (SE) loss of the proposed transceiver and RIS design is derived by considering the limited CSI quantization. To minimize the SE loss, a bit partitioning algorithm that splits the limited number of bits to quantize the CSI is developed.} Extensive numerical results show that the channel customization-based transceiver with reduced CSI can achieve satisfactory performance compared with the optimal transceiver with full CSI. Given the limited number of feedback bits, the bit partitioning algorithm can minimize the SE loss by adaptively allocating bits to quantize the channel parameters.

\end{abstract}

\section{Introduction}

Reviewing the history of wireless communication, the system design techniques usually emerge from the transmitter (Tx) and receiver (Rx) sides because the channel itself is modeled as an uncontrollable exogenous entity \cite{Basar}. Driven by the ever-increasing data rates, future communication systems are envisioned to leverage massive multiple-input-multiple-output (MIMO), millimeter wave (mmWave), and cell-free massive MIMO, and incubate revolutionary technologies, such as integrated sensing and communication, wireless artificial intelligence, and reconfigurable intelligent surfaces (RISs) \cite{road-6G}--\cite{cell-free-m}. Among these technologies, low-cost RISs are envisaged to realize smart radio environments \cite{smart-radio} by manipulating the electromagnetic waves in the channel. Motivated by the limited customization of the radio channel and the substantial development of metamaterials over the past decades \cite{meta}, RISs, which provide one more degree-of-freedom for optimizing wireless communications systems, have recently received widespread attention.\par

As a cost-effective solution for future wireless communication, RISs integrated in various promising applications have been investigated in recent studies \cite{S-CSI-Han}--\cite{App-cell-free}. In a single-user scenario, the downlink performance of RIS-assisted wireless communication was studied in \cite{S-CSI-Han} by deriving a tight upper bound on the ergodic spectral efficiency (SE), when the short-term and long-term channel state information (CSI) are exploited to design the beamforming vector at the base station and the phase shifters at the RIS, respectively. In \cite{app-Wu}, a RIS was introduced in a single-cell wireless system, and passive beamforming was first proposed by adjusting the phase shifters of the surface to minimize the total transmit power. Considering that the RISs consume less energy than the conventional amplify-and-forward relays, the energy efficiency maximization problem was investigated in \cite{App-Huang}, where two effective algorithms that jointly design the transmitting power allocation and the phase shifters of RIS were proposed to tackle the non-convex optimization problem. A sparse array of subsurface (SAoS) deployment that splits a large piece of RIS into small tiles was proposed in \cite{App-DCN} to enhance the coverage of mmWave communication networks. Considering the visible region of the RIS, \cite{App-GC} showed that when the size and distance of RIS tiles are properly optimized, the SAoS can achieve higher SE than the compact deployment with the same number of elements. RISs have been introduced to achieve specific optimization objectives in diverse application scenarios, such as non-orthogonal multiple access  \cite{App-NOMA-Liu}, spatial modulation \cite{App-Spatial},  cognitive radio \cite{App-CR-Liang}, unmanned aerial vehicles \cite{App-UAV}, simultaneous wireless information and power transfer \cite{App-SWIPT}, and cell-free massive MIMO \cite{App-cell-free}. The success of RIS applications relies on the availability of CSI. Channel estimation is indispensable for most wireless systems, and channel feedback is an inherent requirement in frequency division duplexing (FDD) to acquire CSI.\par

In RIS-assisted wireless systems, channel estimation is a challenging task due to the large number of RIS elements that are passive and lack baseband processing capabilities. Numerous studies have addressed the channel estimation problem in RIS-assisted systems. Utilizing the sparse characteristics of mmWave channels, several research teams have attempted to reconstruct the channel by extracting the channel multipath parameters with fibers-missing tensor completion tools \cite{estimation-Lin}, atomic norm minimization \cite{estimation-He}, compressive sensing \cite{estimation-Chen}, and the Newtonized orthogonal matching pursuit algorithm \cite{estimation-NOMP}.  \textcolor[rgb]{0,0,0}{Utilizing the RIS configuration encoding \cite{RIS-coding} to separate signal components from different RISs, the channel estimation algorithms  \cite{estimation-Lin}--\cite{estimation-NOMP} developed for single-RIS-assisted systems can be applied to the multi-RIS-assisted counterpart.} The authors in \cite{estimation-Alwazani} developed an optimal channel estimation protocol utilizing the Bayesian technique to tackle the channel estimation problem for a distributed RIS-assisted MIMO system. Four channel estimation algorithms were proposed in \cite{estimation-Choi} to reduce the training overhead for channel estimation, where the effective channel parameters were estimated at the base station (BS) to reconstruct the cascaded mobile station (MS)--RIS--BS channel. \textcolor[rgb]{0,0,0}{Considering the existence of estimation error, the imperfect CSI obtained from channel estimation was taken into account to design robust and secure transmission for RIS-assisted systems in \cite{estimation-imp}.} More studies on channel estimation for RIS-assisted systems can be found in \cite{estimation-pan} and \cite{estimation-Zheng}.\par

Although channel estimation has received intensive attention, channel feedback for RIS-assisted FDD systems is still in its infancy and should be explored because the RIS is an integral component of the channel itself. In single-user scenarios, a novel cascaded codebook and a bit partitioning strategy were developed in \cite{bit_par} to adaptively quantize and feed back the CSI that has been divided into two parts by the RIS. This process was performed to reduce the rate loss caused by the limited feedback ability of the MS. In multi-user systems assisted by an RIS, overhead reduction feedback schemes exploiting the single-structured sparsity of BS--RIS--MS cascaded channel and the specific triple-structured sparsity of the beamspace cascaded channel were proposed in \cite{feedback-Shen} and \cite{feedback-Shi}, respectively, considering that different MSs share the same sparse BS--RIS channel but have their respective RIS--MS channels. Moreover, \cite{feedback-Prasad} proposed to feed back the signal strength of the intended receiver to reconstruct the channel at the BS rather than feeding back the CSI directly. The proposed channel reconstruction formulation was addressed by efficient proximal distance algorithms and simultaneously exploiting the low-rank property and sparse beamspace representation of the unknown effective channel. The feedback for RIS configurations was investigated in \cite{feedback-Kim} and \cite{feedback-Yu}. A novel codebook-based protocol was proposed in \cite{feedback-Kim} to realize adaptive RIS control with limited feedback considering the angle-dependent behavior \cite{angle-dependent} of the RIS, where the codebook is a set of capacitance values for the RIS configuration. The feedback of the quantized phase shifters was studied in \cite{feedback-Yu}, where a convolutional autoencoder-based scheme was proposed to compress the quantized phase shifter at the Rx side and reconstructed on the RIS side. All the aforementioned studies have focused on channel feedback in single-RIS-assisted systems. To the best of our knowledge, no existing studies have investigated the feedback problem in the multi-RIS-assisted counterpart. In particular, how the customization capability of RISs affects the channel feedback has not been considered.\par

In multi-RIS-assisted sub-6 GHz FDD systems, the multiple high-dimensional individual channels and abundant propagation paths result in prohibitive feedback overhead. Motivated by the potential of RISs in customizing radio channels, we study feedback overhead reduction by configuring RISs with limited CSI to customize the wireless channel. The SE loss minimization under limited quantization is investigated via a bit partitioning strategy. The contributions of our work are summarized as follows.\par
\begin{itemize}
  \item \emph{An efficient channel customization scheme for reducing transceiver complexity and feedback overhead under sub-6 GHz systems}. Different from \cite{channel_cus}, which aims to transform the spatially sparse mmWave channel into a well-conditioned channel that has the minimum truncated condition number, our study proposes to tailor the rich scattering environment in sub-$6$ GHz FDD systems for reducing the feedback overhead. Utilizing the statistical CSI, the composite channel can be customized in the form of singular value decomposition (SVD) by path selection and phase shifters' design. \textcolor[rgb]{0,0,0}{On the basis of channel customization, the optimal SVD transceiver can be easily designed without a matrix decomposition and feeding back the complete channel parameters, which is necessary for existing feedback schemes.} Thus, the computational complexity and feedback overhead are reduced.
  \item \emph{Efficient channel feedback scheme with limited quantization.} We analyze the theoretical SE loss incurred by the limited quantization of the statistical CSI and develop a bit partitioning strategy to reduce it. Although the limited singular values and \textcolor[rgb]{0,0,0}{statistical CSI} are required to be fed back for designing the channel customization-based SVD transceiver, we properly set the size of RISs so that the equal power allocation can be asymptotically equivalent to the optimal water-filling algorithm. This setting normalizes the effect of singular values and further reduces the feedback overhead at the cost of an acceptable performance loss. On this basis, the SE loss caused by the limited quantization of \textcolor[rgb]{0,0,0}{the statistical CSI} is investigated. A bit partitioning strategy is proposed to split the limited feedback bits for separately quantizing the \textcolor[rgb]{0,0,0}{statistical CSI} and minimizing the SE loss.
  \item \emph{Comprehensive simulations combined with insightful discussions about the channel customization scheme.} Numerical simulations are conducted to show the effectiveness of the proposed channel customization and bit partitioning in reducing the feedback overhead and SE loss, respectively. The rich-scattering channel of the multi-RIS-assisted system can be reshaped into a full rank channel that is dominated by sparse orthogonal paths by using the proposed channel customization. Utilizing the perfect limited CSI of the orthogonal paths, the proposed transceiver achieves comparable performance with the optimal SVD transceiver. When limited quantization is considered for feedback, the SE loss is verified to be efficiently reduced with the proposed bit partitioning algorithm.

\end{itemize}\par

The remainder of this paper is organized as follows: Section \ref{sec:2} presents the underlying system model. Sections \ref{sec:3} and \ref{sec:4} investigate the channel customization-based feedback reduction and the ergodic SE loss incurred by limited quantization, respectively. Section \ref{sec:5} discusses the numerical results. Section \ref{sec:6} provides our most important conclusions.\par

\emph{Notations}: A vector and a matrix are denoted by the lowercase and uppercase of a letter, respectively; $|\cdot|$ and $\|\cdot\|_{F}$ are used to indicate the absolute value and Euclidean norm, respectively. Superscripts $(\cdot)^T$ and $(\cdot)^H$ denote the transpose and conjugated-transpose operations, respectively; $\lceil\cdot\rceil$ is the integer ceiling; ${\mathbb C}^{a\times b}$ denotes the set of complex $a\times b$ matrix; ${\mathbb Z}^{+}$ represents the set of positive
numbers; ${\mathbb E}\{\cdot\}$ calculates the statistical expectation. \textcolor[rgb]{0,0,0}{The notation ${\rm blkdiag}\{{\bf X}_1,{\bf X}_2,\ldots,{\bf X}_N\}$ represents a block diagonal matrix with matrices ${\bf X}_i$, $i=1,\ldots,N$}, and ${\rm diag}(a_1,a_2,\ldots,a_N)$ indicates a diagonal matrix with diagonal elements $a_i$, $i=1,\ldots,N$.

\section{System Model}\label{sec:2}
\begin{figure}[!t]
\centering
\includegraphics[width=0.5\textwidth]{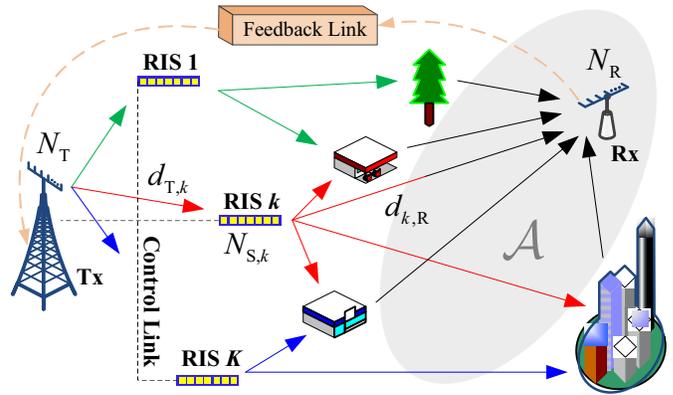}
\caption{RIS-assisted sub-6 GHz FDD system where the downlink CSI is fed back via the feedback link.}
\label{Fig.system_model} 
\end{figure}

We consider a multiple RIS-assisted sub-6 GHz FDD MIMO system, as shown in Fig. \ref{Fig.system_model}, where a blind coverage area $\mathcal A$ appears, when some skyscrapers that obstruct the transmission links from the Tx to the intended area exist. \textcolor[rgb]{0,0,0}{In the blind coverage area caused by blockage, the Rx loses the direct transmission link from the Tx.} Then, $K$ RISs are coated on the facades of of skyscrapers to establish cascaded reflection/refraction links that enable data transmission between the Tx and the Rx rather than setting up another Tx to compensate for the blind coverage area. \textcolor[rgb]{0,0,0}{For the downlink transmission in FDD systems, the downlink CSI is estimated at the Rx and then fed back to the Tx via the feedback link that is independent of the transmission link.} When channel feedback is completed, the Tx uses the quantized downlink CSI to design the precoder and configure the RIS through the control link. For clear and concise description, the Tx, RIS, and Rx are equipped with uniform linear arrays\footnote{This work can be easily extended to systems with uniform planar arrays (UPAs), as will be discussed in the following.} (ULAs) with the inter-element spacing being half wavelength. The number of antennas at the Tx and Rx are denoted by $N_{\rm T}$ and $N_{\rm R}$, respectively. The RISs should be equipped with a sufficient number of elements to combat the diverse path loss among different transmission links built by RISs. Following \cite{channel_cus}, we denote the number of elements in RIS $k$ as $N_{{\rm S},k}$.\par

\subsection{Channel Model}
Given that the direct transmission link from the Tx to the Rx is severely blocked, the channel between the Tx and Rx consists of $K$ components provided by RISs as \cite{cascaded_model}
\begin{equation}\label{eq:H}
  {\bf{H}} = \sum\limits_{k = 1}^K {{\rho _k}{{\bf{H}}_{k,{\rm{R}}}}{{\bf{\Gamma }}_k}{{\bf{H}}_{{\rm{T}},k}}},
\end{equation}
where ${{\bf{\Gamma }}_k} = {\rm{diag}}( {{e^{j{\omega _{k,1}}}}, \ldots ,{e^{j{\omega _{k,{N_{{\rm{S}},k}}}}}}} ) \in {{\mathbb C}^{{N_{{\rm{S}},k}} \times {N_{{\rm{S}},k}}}}$ is the phase shifter response of RIS $k$, ${\rho _k} = \frac{\lambda }{{4\pi {d_{{\rm T},k}}}}\frac{\lambda }{{4\pi {d_{k,{\rm{R}}}}}}$ is the path loss\footnote{\textcolor[rgb]{0,0,0}{This model shows that the Tx--RIS $k$--Rx channel has a path loss that is the product of the path losses of the Tx--RIS subchannel and the RIS--Rx subchannel, which has been theoretically proved in \cite[Eq. (6)]{W. Tang_1} and \cite[Eq. (18)]{Emil} and verified by the experimental results in \cite{W. Tang_1}. Due to this severe path loss cascaded by the RIS, we only consider the transmission link cascaded by one RIS and ignore links cascaded by more than one RIS.}} for the Tx--RIS $k$--Rx link with $\lambda$ being the wavelength, and $d_{{\rm T},k}$ and $d_{k,{\rm{R}}}$ being the distances from the Tx and Rx to the RIS $k$, respectively. In \eqref{eq:H}, ${{\bf{H}}_{{\rm{T}},k}}\in {{\mathbb C}^{{N_{{\rm{S}},k}} \times {N_{{\rm{T}}}}}}$ and ${{\bf{H}}_{k,{\rm{R}}}}\in {{\mathbb C}^{{N_{{\rm{R}}}} \times {N_{{\rm{S}},k}}}}$ denote the Tx--RIS $k$ and RIS $k$--Rx channel, respectively. \textcolor[rgb]{0,0,0}{By extending the channel estimation studies in \cite{estimation-Lin}--\cite{estimation-NOMP} for multi-RIS-assisted systems or applying the channel estimation algorithm in \cite{estimation-Alwazani}, ${{\bf{H}}_{{\rm{T}},k}}$ and ${{\bf{H}}_{k,{\rm{R}}}}$ are assumed to be perfectly estimated at the Rx. Although channel estimation error is unavoidable, it will be mixed together with quantization error when the limited channel feedback is considered. We begin our study with limited feedback and save the research on joint channel estimation and limited feedback as our future work. }\par
In a typical system deployment, where the Tx and RISs are installed at high-rise buildings to guarantee stable transmission, a line of sight (LoS) link exists in the Tx--RIS $k$ channel. \textcolor[rgb]{0,0,0}{Therefore, considering the multipath environment, we introduce a deterministic component to the statistical multipath model, that is developed and verified by measurements in \cite{Saleh}, to describe the Tx--RIS $k$ channel as
\begin{equation}\label{eq:H_T,k}
	\begin{aligned}
		{{\bf{H}}_{{\rm{T}},k}} = \sqrt {{N_{\rm{T}}}{N_{{\rm{S,}}k}}} &\left( {\sqrt {\frac{{{\kappa _{{\rm{T}},k}}}}{{{\kappa _{{\rm{T}},k}} + 1}}} {{\bf{a}}_{{\rm{S,}}k}}\left( {\Theta _{{\rm{T}},k,0}^{\rm{A}}} \right){\bf{a}}_{\rm{T}}^H\left( {\Theta _{{\rm{T}},k,0}^{\rm{D}}} \right)} \right.\\
		&\left. { + \sqrt {\frac{1}{{{\kappa _{{\rm{T}},k}} + 1}}} {\bf{H}}_{{\rm{T}},k}^{{\rm{NLoS}}}} \right),
	\end{aligned}
\end{equation}
where ${\kappa _{{\rm{T}},k}}$ is the Rician factor,} and ${{\bf{a}}_{{\rm{S,}}k}}(\cdot)$ and ${\bf{a}}_{\rm{T}}(\cdot)$ are the array response vectors determined by the \textcolor[rgb]{0,0,0}{directional parameters} ${\Theta _{{\rm{T}},k,0}^{\rm{A}}}=\pi \cos{\theta _{{\rm{T}},k,0}^{\rm{A}}}$ at the RIS $k$ and ${\Theta _{{\rm{T}},k,0}^{\rm{D}}}=\pi \cos{\theta _{{\rm{T}},k,0}^{\rm{D}}}$ at the Tx, respectively, where ${\theta _{{\rm{T}},k,0}^{\rm{A}}}$ is the angle-of-arrival (AoA), and ${\theta _{{\rm{T}},k,0}^{\rm{D}}}$ is the angle-of-departure (AoD). \textcolor[rgb]{0,0,0}{The non-LoS (NLoS) component of ${{\bf{H}}_{{\rm{T}},k}}$ can be expressed as the superposition of scattered wavefronts, that is, ${\bf{H}}_{{\rm{T}},k}^{{\rm{NLoS}}} = 1/\sqrt {{L_{{\rm{T}},k}}} \sum\nolimits_{l = 1}^{{L_{{\rm{T}},k}}} {{\beta _{{\rm{T}},k,l}}{{\bf{a}}_{{\rm{S}},k}}\left( {\Theta _{{\rm{T}},k,l}^{\rm{A}}} \right){\bf{a}}_{\rm{T}}^H\left( {\Theta _{{\rm{T}},k,l}^{\rm{D}}} \right)} $, where $L_{{\rm T},k}$ is the number of NLoS paths and $\beta_{{\rm T},k,l}  \sim \mathcal{CN}( 0,1)$.} The propagation paths for the Tx--RIS $k$ channel are denoted by ${\mathcal L}_{{\rm T},k}=\{0,1,\ldots,L_{{\rm T},k}\}$. Thus, the Tx--RIS $k$ channel can be rewritten as
\begin{equation}\label{eq:re-H_T,k}
  {{\bf{H}}_{{\rm{T}},k}} = \sum\nolimits_{l \in {\mathcal L}_{{\rm T},k}} {{\alpha _{{\rm{T}},k,l}}{{\bf{a}}_{{\rm{S}},k}}\left( {\Theta _{{\rm{T}},k,l}^{\rm{A}}} \right){\bf{a}}_{\rm{T}}^H\left( {\Theta _{{\rm{T}},k,l}^{\rm{D}}} \right)} ,
\end{equation}
where ${\alpha _{{\rm{T}},k,0}} =  \sqrt {\frac{{{N_{\rm{T}}}{N_{{\rm{S,}}k}}}{{\kappa _{{\rm{T}},k}}}}{{{\kappa _{{\rm{T}},k}} + 1}}} $ and ${\alpha _{{\rm{T}},k,l}}=\sqrt {\frac{{{N_{\rm{T}}}{N_{{\rm{S,}}k}}}}{{( {{\kappa _{{\rm{T}},k}} + 1} ){L_{{\rm{T}},k}}}}} {\beta _{{\rm{T}},k,l}}$ ($l>0$) are the effective path gains for the LoS path and the $l$-th NLoS path, respectively. Considering that the Rx is close to the ground, where the propagation environment is more cluttered due to scatterers, the LoS path may not be the dominant component for the RISs--Rx channel, which is different from the Tx--RIS $k$ channel. Therefore, omitting the Rician factor and following \eqref{eq:re-H_T,k} we express the RIS $k$--Rx channel as
\begin{equation}\label{eq:H_R,k}
  {{\bf{H}}_{k,{\rm{R}}}} = \sum\nolimits_{l \in {\mathcal L}_{k,{\rm{R}}}} {{\alpha _{{k,{\rm{R}}},l}}{{\bf{a}}_{\rm{R}}}\left( {\Theta _{{k,{\rm{R}}},l}^{\rm{A}}} \right){\bf{a}}_{{\rm{S,}}k}^H\left( {\Theta _{{k,{\rm{R}}},l}^{\rm{D}}} \right)},
\end{equation}
where ${\mathcal L}_{k,{\rm{R}}}=\{1,\ldots,L_{k,{\rm{R}}}\}$ is the set of propagation paths, $L_{k,{\rm{R}}}$ is the number of propagation paths, ${\bf a}_{\rm R}(\cdot)$ represents the array response vector at the Rx, and ${\alpha _{{k,{\rm{R}}},l}} = \sqrt {\frac{{{N_{\rm{R}}}{N_{{\rm{S,}}k}}}}{{{L_{{k,{\rm{R}}}}}}}} {\beta _{{k,{\rm{R}}},l}}$ with $\beta_{{k,{\rm{R}}},l}  \sim \mathcal{CN}( 0,1)$ is the effective path gain for the $l$-th path in the RIS $k$--Rx channel. Similarly, ${\Theta _{{k,{\rm{R}}},l}^{\rm{A}}}=\pi \cos{\theta _{{k,{\rm{R}}},l}^{\rm{A}}}$ and ${\Theta _{{k,{\rm{R}}},l}^{\rm{D}}} = \pi \cos{\theta _{{k,{\rm{R}}},l}^{\rm{D}}}$ are the \textcolor[rgb]{0,0,0}{directional parameters} with ${\theta _{{k,{\rm{R}}},l}^{\rm{A}}}$ and ${\theta _{{k,{\rm{R}}},l}^{\rm{D}}}$ being the AoA and AoD at the RISs--Rx channel. In \eqref{eq:H_T,k}--\eqref{eq:H_R,k}, the array response vectors of the ULA\footnote{The array response vector of an UPA can be generated by the array response vectors of the vertical and horizontal ULA. Thus, the UPA scenario will introduce another \textcolor[rgb]{0,0,0}{directional parameter} to be quantized for channel feedback. However, the resolution of \textcolor[rgb]{0,0,0}{directional parameters} of an UPA is lower than that of an ULA when the number of antenna elements are the same, which means fewer quantization bits are required for each parameter. Note that the channel customization with UPA has been investigated in spatially sparse mmWave systems in \cite{channel_cus}.} at the Tx, RISs, and the Rx can be uniformly expressed as
\begin{equation}\label{eq:ARV}
  {\bf a}_{X}(Y) = \frac{1}{\sqrt{N_{X}}}{\left[1, e^{jY},\ldots,e^{j(N_{X}-1)Y}\right]^T},
\end{equation}
where $X\in \{\{{\rm T}\},\{{\rm R}\},\{{\rm S},k\}\}$, and $Y$ is the corresponding \textcolor[rgb]{0,0,0}{directional parameter}.\par

\subsection{Downlink SE}
In the downlink transmission, the received signal at the Rx can be expressed as
\begin{equation}\label{eq:y}
  {\bf{y}} = {{\bf{W}}^H}{\bf{HFs}} + {{\bf{W}}^H}{\bf{n}},
\end{equation}
where ${\bf{s}} = {\left[ {{s_1},{s_2}, \ldots ,{s_{{N_{\rm{R}}}}}} \right]^T} \in {\mathbb C}^{N_{\rm R}\times 1}$ satisfying ${\mathbb E}\{{\bf s}{\bf s}^H\}={\bf I}_{{N_{\rm R}}\times{N_{\rm R}}}$ is the transmitted signal from the Tx; ${\bf n}\in {\mathbb C}^{{N_{\rm R}}\times 1}\sim {\mathcal{CN}}(0,\sigma^2 {\bf I}_{{N_{\rm R}}\times{N_{\rm R}}})$ is the additive white Gaussian noise with noise power $\sigma^2$;  ${\bf F}\in{\mathbb C}^{{N_{\rm T}}\times{N_{\rm R}}}$ and ${\bf W}\in{\mathbb C}^{{N_{\rm R}}\times{N_{\rm R}}}$ are the precoder at the Tx and the combiner at the Rx, respectively. The power constraint is $\left\| {{\bf{Fs}}} \right\|_F^2 \le E$, where $E$ is the total power at the Tx. Given the received signal, the downlink SE can be expressed as
\begin{equation}\label{eq:R}
  R = {\log _2}\det \left( {{\bf{I}}_{{N_{\rm R}}\times{N_{\rm R}}} + \frac{1}{{{\sigma ^2}}}{{\bf{W}}^H}{\bf{HF}}{{\bf{F}}^H}{{\bf{H}}^H}{\bf{W}}} \right).
\end{equation}
In conventional wireless communication systems, when the channel $\bf H$ is fixed, the optimal precoder and combiner that maximize the SE can be obtained by the SVD of $\bf H$. The SVD of $\bf H$ with $N_{\rm T}>N_{\rm R}$ is expressed as
\begin{equation}\label{eq:SVD}
   {\bf{H}} = {{\bf U}}\left[ {{\bf{\Lambda }},{{\bf{0}}_{ {{N_{\rm{R}}}}  \times \left( {{N_{\rm{T}}} -{N_{\rm{R}}}} \right)}}} \right]{\hat{\bf V}}^H  = {\bf{U\Lambda }}{{\bf{V}}^H},
\end{equation}
where ${{\bf U}}\in {\mathbb C}^{{N_{\rm R}}\times{N_{\rm R}}}$ and ${\hat{\bf V}}\in {\mathbb C}^{{N_{\rm T}}\times{N_{\rm T}}}$ are unitary matrices, and ${\bf{\Lambda }}={\rm diag}(\sqrt{\lambda_1},\ldots,\sqrt{\lambda_{N_{\rm R}}})$ is a ${{N_{\rm R}}\times{N_{\rm R}}}$ diagonal matrix with $\sqrt{\lambda_{i}}$ being the $i$-th largest singular value. Note that $\bf V$ is formed by the first $N_{\rm R}$ columns of $\hat{\bf V}$. With the SVD of $\bf H$, the optimal precoder and combiner can be expressed as
\begin{equation}\label{eq:optimal_F_W}
\begin{aligned}
  {\bf{F}} &= {\bf{V}}{{\bf{P}}^{1/2}},\\
  {\bf{W}} &= {\bf{U}},
\end{aligned}
\end{equation}
where ${\bf P}={\rm diag}({ p}_1,\ldots,{ p}_{N_{\rm R}})$ is the water-filling power allocation matrix, whose $i$-th diagonal element is expressed as
\begin{equation}\label{eq:water_p_i}
  {p_i} = \max{\left(0, {\mu  - \frac{{\sigma ^2}}{{\lambda _i}}} \right) },
\end{equation}
where $\mu$ is the water lever that satisfies the power constraint $\sum\nolimits_{i = 1}^{N_{\rm R}} {{p_i}}  = E$. With the optimal precoder and combiner, the SE can be maximized as
\begin{equation}\label{eq:R_max}
  R=\sum\limits_{i=1}^{N_{\rm R}}{\log_2{\left(1+\frac{\lambda_i p_i}{\sigma^2}\right)}}.
\end{equation}\par

The acquisition of $\bf H$ is the precondition for the optimal SVD transceiver in \eqref{eq:optimal_F_W}. In FDD systems, the Tx obtains $\bf H$ through channel feedback. Although channel feedback has been widely investigated in concurrent wireless communication systems \cite{Nihar}, \cite{Ramta_1}, the analysis is different in the RIS space. As observed in \eqref{eq:H}, the phase shifters of RISs are encapsulated in the composite channel $\bf H$. Thus, the first difference is that the Rx cannot feed back $\bf H$ directly to the Tx because the design of $\{{\bf \Gamma}_k\}_{k=1}^{K}$ will accordingly alter the channel. In other words, acquiring the current $\bf H$ from a previous feedback realization, when the phase shifters are to be redesigned, is meaningless. The second difference is that the optimization of RISs for creating a favorable composite channel requires the segmented individual channels $\{ {{{\bf{H}}_{{k,{\rm{R}}}}},{{\bf{H}}_{{\rm{T}},k}}} \}_{k = 1}^K$. However, given that the RISs are planned to be equipped with a massive number of elements, directly feeding back these high-dimensional individual channels is impractical due to the unbearable feedback overhead. On the basis of the geometric channel models \eqref{eq:re-H_T,k} and \eqref{eq:H_R,k}, $\{ {{{\bf{H}}_{{k,{\rm{R}}}}},{{\bf{H}}_{{\rm{T}},k}}} \}_{k = 1}^K$ can be reconstructed at the Tx via the channel parameters
\begin{equation}\label{eq:para_T}
  \left\{ {\left( {{\alpha _{{\rm{T}},k,l}},\Theta _{{\rm{T}},k,l}^{\rm{A}},\Theta _{{\rm{T}},k,l}^{\rm{D}}} \right),l \in {{\mathcal L}_{{\rm{T}},k}}} \right\}_{k = 1}^K,
\end{equation}
\begin{equation}\label{eq:para_R}
  \left\{ {\left( {{\alpha _{{k,{\rm{R}}},i}},\Theta _{{k,{\rm{R}}},i}^{\rm{A}},\Theta _{{k,{\rm{R}}},i}^{\rm{D}}} \right),i \in {{\mathcal L}_{{k,{\rm{R}}}}}} \right\}_{k = 1}^K.
\end{equation}
Considering that the total number of channel parameters ${\sum\nolimits_{k=1}^{K}}{3(1+L_{{\rm T},{k}}+L_{{k,{\rm{R}}}})}$ is less than that of channel matrices, feeding back the channel parameters rather than the channel matrices in RIS-assisted systems is meaningful. Although this scheme can reduce the feedback overhead, the substantial number of scatterers in the electromagnetic propagation environment of the sub-6 GHz systems create a high number of paths. These channel parameters cannot be perfectly fed back to the Tx because the feedback ability of the Rx is limited and parameter quantization is necessary.\par

Given the above challenges, finite CSI with limited quantization should be utilized to approach the SVD transceiver in RIS-assisted FDD systems. To achieve this goal, we propose to customize the composite channel with limited CSI in the forthcoming sections so that the feedback overhead can be greatly reduced. The SE loss resulting from limited quantization is evaluated.

\section{Channel Customization with Limited CSI}\label{sec:3}
In this section, the structure of the composite channel is first analyzed for channel customization-oriented RIS design with limited CSI. We then proceed to customize the channel in the form of SVD. The SVD transceiver can be easily designed without a matrix decomposition and feeding back the complete channel parameters by harnessing the channel customization.\par
Substituting \eqref{eq:re-H_T,k} and \eqref{eq:H_R,k} into \eqref{eq:H}, the composite channel can be rewritten as
\begin{equation}\label{eq:H-1}
	\begin{aligned}
		{\bf{H}} = \sum\limits_{k = 1}^K \sum\limits_{{l_k} = 1}^{{L_{{k,{\rm{R}}}}}} \sum\limits_{{j_k} = 0}^{{L_{{\rm{T}},k}}} &{\rho _k}{\alpha _{{k,{\rm{R}}},{l_k}}}{\alpha _{{\rm{T}},k,{j_k}}}{{\bf{a}}_{\rm{R}}}\left( {\Theta _{{k,{\rm{R}}},{l_k}}^{\rm{A}}} \right){\bf{a}}_{{\rm{S,}}k}^H\left( {\Theta _{{k,{\rm{R}}},{l_k}}^{\rm{D}}} \right)\\
		&\times{{\bf{\Gamma }}_k}{{\bf{a}}_{{\rm{S}},k}}\left( {\Theta _{{\rm{T}},k,{j_k}}^{\rm{A}}} \right){\bf{a}}_{\rm{T}}^H\left( {\Theta _{{\rm{T}},k,{j_k}}^{\rm{D}}} \right)   .
	\end{aligned}
\end{equation}
We denote 
\begin{equation}
	{\xi _{k,{l_k},{j_k}}} \buildrel \Delta \over = {\rho _k}{\alpha _{{k,{\rm{R}}},{l_k}}}{\alpha _{{\rm{T}},k,{j_k}}}{\bf{a}}_{{\rm{S,}}k}^H\left( {\Theta _{{k,{\rm{R}}},{l_k}}^{\rm{D}}} \right){{\bf{\Gamma }}_k}{{\bf{a}}_{{\rm{S}},k}}\left( {\Theta _{{\rm{T}},k,{j_k}}^{\rm{A}}} \right)
\end{equation}
as the effective path gain of the cascaded path from the $j_k$-th path of the TX--RIS $k$ channel to the $l_k$-th path of the RIS $k$--Rx channel. Equation \eqref{eq:H-1} can be expressed as
\begin{equation}\label{eq:H-2}
\begin{aligned}
  {\bf{H}} &= \sum\limits_{k = 1}^K {\sum\limits_{{l_k} = 1}^{{L_{{k,{\rm{R}}}}}} {\sum\limits_{{j_k} = 0}^{{L_{{\rm{T}},k}}} {{\xi _{k,{l_k},{j_k}}}{{\bf{a}}_{\rm{R}}}\left( {\Theta _{{k,{\rm{R}}},{l_k}}^{\rm{A}}} \right){\bf{a}}_{\rm{T}}^H\left( {\Theta _{{\rm{T}},k,{j_k}}^{\rm{D}}} \right)} } }\\
  &= \sum\limits_{k = 1}^K {{{\bf{A}}_{{k,{\rm{R}}}}}{{\bf{\Xi }}_k}{\bf{A}}_{{\rm{T}},k}^H} ,
\end{aligned}
\end{equation}
where ${{\bf{A}}_{{k,{\rm{R}}}}} = [ {{\bf{a}}_{\rm{R}}}( {\Theta _{{k,{\rm{R}}},1}^{\rm{A}}} ), \ldots ,{{\bf{a}}_{\rm{R}}}( {\Theta _{{k,{\rm{R}}},{L_{{k,{\rm{R}}}}}}^{\rm{A}}} ) ] \in {\mathbb C}^{{N_{\rm R}}\times {L_{{k,{\rm{R}}}}}}$ and ${{\bf{A}}_{{\rm{T}},k}} = [ {{{\bf{a}}_{\rm{T}}}( {\Theta _{{\rm{T}},k,0}^{\rm{D}}} ), \ldots ,{{\bf{a}}_{\rm{R}}}( {\Theta _{{\rm{T}},k,{L_{{\rm{T}},k}}}^{\rm{D}}} )} ]\in {\mathbb C}^{N_{\rm T}\times (L_{{\rm T},k}+1)}$ are the array response matrices at the Rx and Tx for the Tx--RIS $k$--Rx channel, respectively; ${{{\bf{\Xi }}_k}} \in {\mathbb C}^{L_{{k,{\rm{R}}}}\times (L_{{\rm T},k}+1)}$ is the effective cascaded path gain matrix for the Tx--RIS $k$--Rx channel with the element in the $l_k$ row and the $j_k$ column being ${( {{{\bf{\Xi }}_k}} )_{{l_k},{j_k}}} = {\xi _{k,{l_k},{j_k}-1}}$. Combining all the channel components provided by $K$ RISs, the channel can be rewritten as
\begin{equation}\label{eq:H-3}
  {\bf{H}} = {{\bf{A}}_{\rm{R}}}{\bf{\Xi A}}_{\rm{T}}^H,
\end{equation}
where \textcolor[rgb]{0,0,0}{${{\bf{A}}_{\rm{R}}} = [ {{{\bf{A}}_{1,{\rm{R}}}}, \ldots ,{{\bf{A}}_{{K,{\rm{R}}}}}} ]\in {\mathbb C}^{{N_{\rm R}}\times L_{\rm R}}$ with $L_{\rm R}=\sum\nolimits_{k=1}^{K}{L_{{k,{\rm{R}}}}}$, ${{\bf{A}}_{\rm{T}}} = [ {{\bf{A}}_{{\rm{T}},1}}, \ldots ,{{\bf{A}}_{{\rm{T}},K}} ]\in {\mathbb C}^{{N_{\rm T}}\times L_{\rm T}}$ with $L_{\rm T}=\sum\nolimits_{k=1}^{K}({L_{{\rm T},k}}+1)$, and ${\bf{\Xi }} = {\rm{blkdiag}}( {{{\bf{\Xi }}_1}, \ldots ,{{\bf{\Xi }}_K}} )\in {\mathbb C}^{L_{\rm R}\times L_{\rm T}}$.}\par
For the raw composite channel, where the RISs are not designed, that is, ${\bf \Gamma}_k={\bf I}_{{ N_{{\rm S},k} }\times{N_{{\rm S},k}}}$, $\forall k \in {\mathcal K}$, the effective cascaded path gain can be reduced to
\begin{equation}\label{eq:xi-I}
  {\xi _{k,{l_k},{j_k}}} = {\rho _k}{\alpha _{{k,{\rm{R}}},{l_k}}}{\alpha _{{\rm{T}},k,{j_k}}}{\bf{a}}_{{\rm{S,}}k}^H\left( {\Theta _{{k,{\rm{R}}},{l_k}}^{\rm{D}}} \right){{\bf{a}}_{{\rm{S}},k}}\left( {\Theta _{{\rm{T}},k,{j_k}}^{\rm{A}}} \right).
\end{equation}
Considering the asymptotic orthogonality of array response vectors in the large array regime \cite{bit_par}, where ${N_{{\rm{S}},k}} \to \infty $, when $\Theta _{{k,{\rm{R}}},{l_k}}^{\rm{D}} \ne \Theta _{{\rm{T}},k,{j_k}}^{\rm{A}}$, we will have
\begin{equation}\label{eq:asymp_orth}
  {\bf{a}}_{{\rm{S,}}k}^H\left( {\Theta _{{k,{\rm{R}}},{l_k}}^{\rm{D}}} \right){{\bf{a}}_{{\rm{S}},k}}\left( {\Theta _{{\rm{T}},k,{j_k}}^{\rm{A}}} \right) \to 0.
\end{equation}
This asymptotic orthogonality implies that if all RISs are not controlled, ${\bf{\Xi }} \to {\bf{0}}$, and the rank of $\bf H$ will be 0. In such a case, the Rx has no sustainable connection with the Tx. To reap the maximal spatial multiplexing gain at the Rx so that $N_{\rm R}$ data streams can be effectively received from the Tx, we will focus on tuning the phase shifters to shape a $N_{\rm R}$-rank channel.\par

\begin{assumption}\label{assump:1}
In all propagation paths between the Tx--RISs channel, a path set $\mathop  \cup \limits_{k = 1}^K {\mathcal L}_{{\rm{T}},k}^\star $ (${\mathcal L}_{{\rm{T}},k}^\star  \subseteq {{\mathcal L}_{{\rm{T}},k}}$ and $\sum\nolimits_{k = 1}^K {| {{\mathcal L}_{{\rm{T}},k}^\star } |}  = {N_{\rm{R}}}$) exists,  where the corresponding array response vectors at the Tx are pair-wisely orthogonal, that is,
\begin{equation}\label{eq:a_T-ortho}
  {\bf{a}}_{\rm{T}}^H\left( {\Theta _{{\rm{T}},{m},i_m}^{\rm{D}}} \right){{\bf{a}}_{\rm{T}}}\left( {\Theta _{{\rm{T}},{n},i_n}^{\rm{D}}} \right) = 0,
\end{equation}
that holds for $i_n, i_m \in \mathop  \cup \limits_{k = 1}^K {\mathcal L}_{{\rm{T}},k}^\star$ and $i_n \ne i_m$. Similarly, another path set $\mathop  \cup \limits_{k = 1}^K {\mathcal L}_{{k,{\rm{R}}}}^\star $ (${\mathcal L}_{{k,{\rm{R}}}}^\star  \subseteq {{\mathcal L}_{{k,{\rm{R}}}}}$ and $ {| {{\mathcal L}_{{k,{\rm{R}}}}^\star } |}  ={| {{\mathcal L}_{{\rm{T}},k}^\star } |}$) exists in the RISs--Rx channels, where the corresponding array response vectors at the Rx are pair-wisely orthogonal.
\end{assumption}\par
On the basis of \emph{Assumption \ref{assump:1}}, the active RISs that contribute at least one orthogonal path are represented as ${{\mathcal K}_ \bot } = \{ {k ,| {{\mathcal L}_{{\rm{T}},k}^\star } | \geq1 } \}$. We then split the composite channel into non-orthogonal and orthogonal parts as
\begin{equation}\label{eq:H-4}
  {\bf{H}} = {{\bf{A}}_{\rm{R}}}{{\bf{\Xi }}_\parallel }{\bf{A}}_{\rm{T}}^H + {{\bf{A}}_{{\rm{R,}} \bot }}{{\bf{\Xi }}_ \bot }{\bf{A}}_{{\rm{T,}} \bot }^H,
\end{equation}
where ${{\bf{\Xi }}_\parallel }$ is the counterpart of ${{\bf{\Xi }} }$ by replacing ${\xi _{k,{l_k},{j_k}}}$ with $0$ for $k \in {{\mathcal K}_ \bot }$, ${l_k} \in {\mathcal L}_{{k,{\rm{R}}}}^\star$, and ${j_k} \in {\mathcal L}_{{\rm{T}},k}^\star$; \textcolor[rgb]{0,0,0}{${{\bf{\Xi }}_ \bot }={\rm blkdiag}\{{{\bf{\Xi }}_{1, \bot }},\ldots,{{\bf{\Xi }}_{|{{\mathcal K}_ \bot }|, \bot }}\} \in {{\mathbb C}^{{N_{\rm{R}}} \times {N_{\rm{R}}}}}$ is a block diagonal matrix with ${( {{{\bf{\Xi }}_{k, \bot }}} )_{{l_k},{j_k}}} = {\xi _{k,{l_k},{j_k}-1}}$ for $k \in {{\mathcal K}_ \bot }$, ${l_k} \in {\mathcal L}_{{k,{\rm{R}}}}^\star$, and ${j_k} \in {\mathcal L}_{{\rm{T}},k}^\star$. In \eqref{eq:H-4}, ${{\bf{A}}_{{\rm{R,}} \bot }} =[{{\bf{A}}_{{1,{\rm{R}}}, \bot }},\ldots,{{\bf{A}}_{{|{{\mathcal K}_ \bot }|,{\rm{R}}}, \bot }}] \in {{\mathbb C}^{{N_{\rm{R}}} \times {N_{\rm{R}}}}}$, where the columns of ${{\bf{A}}_{{k,{\rm{R}}}, \bot }}$ are $\{ {{{\bf{a}}_{\rm{R}}}( {\Theta _{{k,{\rm{R}}},{l_k}}^{\rm{A}}} ),{l_k} \in {\mathcal L}_{{k,{\rm{R}}}}^\star } \}$, and ${{\bf{A}}_{{\rm{T,}} \bot }} =[{{\bf{A}}_{{1,{\rm{T}}}, \bot }},\ldots,{{\bf{A}}_{{|{{\mathcal K}_ \bot }|,{\rm{T}}}, \bot }}] \in {{\mathbb C}^{{N_{\rm{T}}} \times {N_{\rm{R}}}}}$, while the columns of ${{\bf{A}}_{{\rm{T,}}k, \bot }}$ are $\{ {{{\bf{a}}_{\rm{T}}}( {\Theta _{{\rm{T}},k,{j_k}}^{\rm{D}}} ),{j_k} \in {\mathcal L}_{{\rm{T}},k}^\star } \}$.} We point out that $N_{\rm R}$ orthogonal departure paths and $N_{\rm R}$ orthogonal arrival paths are extracted at the Tx and Rx by representing the composite channel as \eqref{eq:H-4}, respectively.\par

When the phase shifters of RISs in ${\mathcal K}_\bot$ are rigorously designed to shape ${{\bf{\Xi }}_{k, \bot }}$ as a diagonal matrix satisfying ${{\bf{\Xi }}_\parallel }\to {\bf 0}$, ${{\bf{A}}_{{\rm{R,}} \bot }}{{\bf{\Xi }}_ \bot }{\bf{A}}_{{\rm{T,}} \bot }^H$ in \eqref{eq:H-4} can be approximated as the SVD of $\bf H$. A full row rank channel is customized, and the SVD transceiver can be easily obtained without a matrix decomposition. Motivated by the above observations, we aim at channel customization by appropriately selecting paths to construct ${{\bf{A}}_{{\rm{R,}} \bot }}$ and ${{\bf{A}}_{{\rm{T,}} \bot }}$, and designing RISs to modify ${{\bf{\Xi }}_{k, \bot }}$ and ${{\bf{\Xi }}_\parallel }$ with limited CSI. The feedback overhead required for channel customization-based transceiver is analyzed.\par
\subsection{Path Selection}\label{sec:3.1}
Considering that the Tx and RISs are stationary, we follow \cite{channel_cus} by assuming that the RISs have been installed at different DFT directions\footnote{It is worth noting that the proposed channel customization is extendable to the scenarios where the RISs are not at the DFT directions of the Tx. In such cases, path selection required for ${\bf A}_{{\rm T},\bot}$ is similar to that for ${\bf A}_{{\rm R},\bot}$.} of the Tx during system deployment, that is,
\begin{equation}\label{eq:RISs-DFT}
  {\Theta _{{\rm{T}},k,0}^{\rm{D}}}\in \left\{\frac{2\pi n}{N_{\rm T}}+\Delta, n=1,2,\ldots,N_{\rm T}\right\}, \forall k,
\end{equation}
where $\Delta$ can take an arbitrary value in $[0,2\pi]$. Under this deployment, all LoS paths of the Tx--RISs channel can be selected as orthogonal paths to satisfy \eqref{eq:a_T-ortho}, indicating ${\mathcal L}_{{\rm{T}},k}^\star  = \{ 0 \},k \in {\mathcal K}$. Deploying RISs at the DFT directions of the Tx simplifies the path selection for constructing ${\bf A}_{{\rm T},\bot}$ and reduces the feedback overhead, as will be discussed in Section \ref{sec:3.3}.\par

Given that ${\bf A}_{{\rm T},\bot}$ can be determined by picking $N_{\rm R}$ LoS paths from the Tx--RISs channel, we first consider path selection for ${\bf A}_{{\rm R},\bot}$. Considering that $| {{\mathcal L}_{{k,{\rm{R}}}}^\star } | = | {{\mathcal L}_{{\rm{T}},k}^\star } | = 1$, only one path will be selected in the RIS $k$--Rx channel. Therefore, constructing ${\bf A}_{{\rm R},\bot}$ with mutually orthogonal columns should individually pick up $N_{\rm R}$ paths from $N_{\rm R}$ RIS--Rx channels, which can be straightforwardly formulated as\textcolor[rgb]{0,0,0}{
\begin{equation}\label{eq:A_R_or}
  \begin{aligned}
{{\bf{A}}_{{\rm{R,}} \bot }} &= \mathop {\arg \min }\limits_{\bf{A}} \left\| {{{\bf{A}}^H}{\bf{A}} - {\bf{I}}_{{N_{\rm R}}\times{N_{\rm R}}}} \right\|_F^2\\
{\rm s. t. }\quad&{\left[ {\bf{A}} \right]_{:,i}} = {{\bf{a}}_{\rm{R}}}\left( {\Theta _{{k_i,{\rm{R}}},{l_{k_i}}}^{\rm{A}}} \right),\; i=1,\ldots,N_{\rm R} \\
&k_i \in {{\mathcal K}_ \bot },\;{{\mathcal K}_ \bot } \subseteq {\mathcal K},\; \left| {{{\mathcal K}_ \bot }} \right| = {N_{\rm{R}}}\\
&{l_{k_i}} \in {{\mathcal L}^{\star}_{{{k_i},{\rm{R}}}}},\; {{\mathcal L}^{\star}_{{{k_i},{\rm{R}}}}} \subseteq {{\mathcal L}_{{{k_i},{\rm{R}}}}},\; \left|{{\mathcal L}^{\star}_{{{k_i},{\rm{R}}}}}\right| = 1.
\end{aligned}
\end{equation}}
The optimal solution for the combinational problem \eqref{eq:A_R_or} requires an exhaustive search, where its computational complexity is
\textcolor[rgb]{0,0,0}{\begin{equation}\label{eq:search-time}
  \sum\limits_{n = 1}^{{Z_{K,{N_{\rm{R}}}}}} {\prod\limits_{k = 1}^K {L_{{k,{\rm{R}}}}^{{I_{n,k}}}} } ,
\end{equation}}
where the coefficient of combinatorial optimization is defined as
\begin{equation}\label{eq:Z_K_N_R}
  {Z_{K,{N_{\rm{R}}}}} = \left( {\begin{array}{*{20}{c}}
K\\
{{N_{\rm{R}}}}
\end{array}} \right) = \frac{{K!}}{{\left( {K - {N_{\rm{R}}}} \right)!{N_{\rm{R}}}!}}.
\end{equation}
In \eqref{eq:search-time}, ${I_{n,k}} \in \{ {1,0} \}$ indicates whether RIS $k$ is selected in the $n$-th combination or not, which satisfies $\sum\nolimits_{k = 1}^K {{I_{n,k}}}  = {N_{\rm{R}}}$ (only $N_{\rm R}$ RISs will be selected for the $n$-th combination) and $\sum\nolimits_{n = 1}^{{Z_{K,{N_{\rm{R}}}}}} {{I_{n,k}}}  = {Z_{K - 1,{N_{\rm{R}}} - 1}}$ (RIS $k$ will be selected in ${Z_{K - 1,{N_{\rm{R}}} - 1}}$ combinations).\par

\textcolor[rgb]{0,0,0}{Solving problem \eqref{eq:A_R_or} by exhaustively searching all paths produces ${\bf A}_{{\rm R},\bot}$,} whose columns are approximately orthogonal to each other. However, the search complexity is huge. For example, when $K=6$, $N_{\rm R}=4$, and $\forall L_{{k,{\rm{R}}}}=10$, the search complexity is $1.5\times 10^5$. Weak paths that have small gain ${\alpha _{{k,{\rm{R}}},{l_k}}}$ are likely to be selected, which will decrease the SE. Path pruning is introduced to reduce the search complexity while increasing the SE. Before exhaustively searching for \eqref{eq:A_R_or}, the weak path $l_k$ is omitted from ${\mathcal L}_{{k,{\rm{R}}}}$ when
\begin{equation}\label{eq:path-pruning}
  \frac{\left|{\alpha _{{k,{\rm{R}}},{l_k}}}\right|}{\mathop {\max }\limits_{{n_k} \in {{\mathcal L}_{{k,{\rm{R}}}}}} \left| {{\alpha _{{k,{\rm{R}}},{n_k}}}} \right|}<\tau,
\end{equation}
is satisfied, where $\tau $ is a threshold. \textcolor[rgb]{0,0,0}{Considering that weak paths are not involved in the path selection process, the proposed path pruning also reduces the channel estimation complexity since only strong paths need to be accurately estimated. Although the optimal solution of \eqref{eq:A_R_or} cannot be achieved with path pruning since the searchable space is squeezed, the sub-optimal solution with stronger paths can increase the SE as will be shown in Fig. \ref{Fig.pruning} and the search complexity can be reduced from $1.5\times 10^5$ to $9375$ if half paths are weak and omitted.} Then, ${\bf A}_{{\rm R},\bot}$ and ${\mathcal K}_\bot$ are obtained, and ${\bf A}_{{\rm T},\bot}$ can be determined by selecting the LoS path that connects the Tx and RIS $k\in {\mathcal K}_\bot$ after path pruning \eqref{eq:path-pruning} and selection \eqref{eq:A_R_or} in the RISs--Rx channel are completed.

\subsection{RIS Design}\label{sec:3.2}
\emph{Assumption \ref{assump:1}} can be realized and the composite channel split in \eqref{eq:H-4} is achieved by utilizing path selection. We now focus on designing the phase shifters to enhance the diagonal element of ${{\bf{\Xi }}_ \bot }$ while maintaining ${{\bf{\Xi }}_\parallel } \to {\bf{0}}$.\par

Given that $| {{\mathcal L}_{{k,{\rm{R}}}}^\star } | = | {{\mathcal L}_{{\rm{T}},k}^\star } | = 1$ for $k\in {\mathcal K}_{\bot}$, ${{\bf{\Xi }}_ \bot }$ is a diagonal matrix, whose element is
\begin{equation}\label{eq:Xi-diag}
   {{\xi _{k,{l_k},0}} = {\rho _k}{\alpha _{{k,{\rm{R}}},{l_k}}}{\alpha _{{\rm{T}},k,0}}{\bf{a}}_{{\rm{S,}}k}^H\left( {\Theta _{{k,{\rm{R}}},{l_k}}^{\rm{D}}} \right){{\bf{\Gamma }}_k}{{\bf{a}}_{{\rm{S}},k}}\left( {\Theta _{{\rm{T}},k,0}^{\rm{A}}} \right) },
\end{equation}
for $k \in {{\mathcal K}_ \bot }$ and $\textcolor[rgb]{0,0,0}{l_k \in {\mathcal L}_{{k,{\rm{R}}}}^\star}$. The phase shifters that maximize ${\xi _{k,{l_k},0}}$ can be expressed as \cite{channel_cus}
\begin{equation}\label{eq:optimal-RIS}
  {\omega _{k,n}} = \left( {n - 1} \right)\left( {\Theta _{{k,{\rm{R}}},{l_k}}^{\rm{D}} - \Theta _{{\rm{T}},k,0}^{\rm{A}}} \right),
\end{equation}
where $n \in \left\{ {1, \ldots ,{N_{{\rm{S}},k}}} \right\}$ and ${l_k} \in {\mathcal L}_{{k,{\rm{R}}}}^\star $. The singular values of ${{\bf{A}}_{{\rm{R,}} \bot }}{{\bf{\Xi }}_ \bot }{\bf{A}}_{{\rm{T,}} \bot }^H$ can be maximized as $\{ {| {\xi _{k,{l_k},0}^\star } | = | {{\rho _k}{\alpha _{{k,{\rm{R}}},{l_k}}}{\alpha _{{\rm{T}},k,0}}} |,k \in {{\mathcal K}_ \bot },\textcolor[rgb]{0,0,0}{l_k \in {\mathcal L}_{{k,{\rm{R}}}}^\star} } \}$ by designing the phase shifters of the RISs in ${\mathcal K}_\bot$ in accordance with \eqref{eq:optimal-RIS}. For the remaining effective cascaded path gains ${\xi _{k,{l_k},{j_k}}}$ ($k \in {{\mathcal K}_ \bot }$, ${l_k} \in {{\mathcal L}_{{k,{\rm{R}}}}}\backslash {\mathcal L}_{{k,{\rm{R}}}}^\star $, and ${j_k} \in {{\mathcal L}_{{\rm{T}},k}}\backslash {\mathcal L}_{{\rm{T}},k}^\star $), the RIS design in \eqref{eq:optimal-RIS} yields
\begin{equation}\label{eq:rest-xi}
	\begin{aligned}
		{\xi _{k,{l_k},{j_k}}} = {\rho _k}{\alpha _{{k,{\rm{R}}},{l_k}}}{\alpha _{{\rm{T}},k,{j_k}}}{\bf{a}}_{{\rm{S,}}k}^H\left( {\Theta _{{k,{\rm{R}}},{l_k}}^{\rm{D}} - \Theta _{{k,{\rm{R}}},l_k^\star }^{\rm{D}}} \right)\\
		\times{{\bf{a}}_{{\rm{S}},k}}\left( {\Theta _{{\rm{T}},k,{j_k}}^{\rm{A}} - \Theta _{{\rm{T}},k,j_k^\star }^{\rm{A}}} \right),
	\end{aligned}
\end{equation}
where $l_k^\star  \in {\mathcal L}_{{k,{\rm{R}}}}^\star $ and $j_k^\star  \in {\mathcal L}_{{\rm{T}},k}^\star $. In accordance with the asymptotic orthogonality of the array response vectors, we have
\begin{equation}\label{eq:asym-ortho-2}
   {\bf{a}}_{{\rm{S,}}k}^H\left( {\Theta _{{k,{\rm{R}}},{l_k}}^{\rm{D}} - \Theta _{{k,{\rm{R}}},l_k^\star }^{\rm{D}}} \right){{\bf{a}}_{{\rm{S}},k}}\left( {\Theta _{{\rm{T}},k,{j_k}}^{\rm{A}} - \Theta _{{\rm{T}},k,j_k^\star }^{\rm{A}}} \right)\to 0,
\end{equation}
when $N_{{\rm S},{k}}\to \infty$ and $ {\Theta _{{k,{\rm{R}}},{l_k}}^{\rm{D}} - \Theta _{{k,{\rm{R}}},l_k^\star }^{\rm{D}}} \neq  {\Theta _{{\rm{T}},k,{j_k}}^{\rm{A}} - \Theta _{{\rm{T}},k,j_k^\star }^{\rm{A}}} $. \textcolor[rgb]{0,0,0}{For RIS $k \in {\mathcal K} \backslash {\mathcal K}_{\bot}$, we keep them undesigned\footnote{\textcolor[rgb]{0,0,0}{In this study, $N_{\rm R}$ RISs are activated to customize a sparse channel for the Rx while the remaining RISs are left undesigned because we consider the single Rx scenario. By developing proper RIS--Rx association and RISs scheduling strategies to select and design all RISs for interference elimination, this work can be extended to multi-Rxs case. }}, that is, ${\bf \Gamma}_k={\bf I}_{{ N_{{\rm S},k} }\times{N_{{\rm S},k}}}$.} Therefore, ${{\bf{\Xi }}_\parallel } \to {\bf{0}}$ still holds with the RIS design \eqref{eq:optimal-RIS}, and the composite channel $\bf H$ can be approximately customized in the form of SVD as
\begin{equation}\label{eq:SVD-appro}
  {\bf{H}} \approx {{\bf{A}}_{{\rm{R,}} \bot }}{{\bf{\Xi }}_ \bot }{\bf{A}}_{{\rm{T,}} \bot }^H.
\end{equation}
The essential channel parameters to accomplish channel customization are
\begin{equation}\label{eq:CC-para}
  \left\{ {\left( {\Theta _{{\rm{T}},k,{j_k}}^{\rm{A}},\Theta _{{k,{\rm{R}}},{l_k}}^{\rm{D}}} \right),k \in {{\mathcal K}_ \bot },{j_k} \in {\mathcal L}_{{\rm{T}},k}^\star ,{l_k} \in {\mathcal L}_{{k,{\rm{R}}}}^\star } \right\}.
\end{equation}

\subsection{Feedback Overhead for the Channel Customization-based SVD Transceiver}\label{sec:3.3}
\textcolor[rgb]{0,0,0}{Compared with the conventional SVD transceiver that requires an SVD process for the composite channel, the channel customization-based SVD transceiver can be obtained without matrix decomposition, that is,
\begin{equation}\label{eq:CC-SVD}
\begin{aligned}
{\bf{F}} &= {{\bf{A}}_{{\rm{T,}} \bot }}{{\bf{P}}^{1/2}},\\
{\bf{W}} &= {{\bf{A}}_{{\rm{R,}} \bot }},
\end{aligned}
\end{equation}
where $\bf P$ is a diagonal matrix, where its non-zero elements are $\{ {{p_k},k \in {{\mathcal K}_ \bot }} \}$, and ${p_k} = \max \big( {\mu  - {{\sigma ^2} \over{{| {\xi _{k,{l_k},0}^\star } |}^2}},0} \big)$ is the water-filling power allocation coefficient.} Considering that the additional water-filling power allocation is determined by the singular values, the amount of CSI required for channel customization-based SVD transceiver is limited, that is,
\begin{equation}\label{eq:SVD-para-3}
  \left\{ {\left( {| {\xi _{k,{l_k},{j_k}}^\star } |},{\Theta _{{\rm{T}},k,{j_k}}^{\rm{A}},\Theta _{{k,{\rm{R}}},{l_k}}^{\rm{D}}} \right),k \in {{\mathcal K}_ \bot },{j_k} \in {\mathcal L}_{{\rm{T}},k}^\star ,{l_k} \in {\mathcal L}_{{k,{\rm{R}}}}^\star } \right\}.
\end{equation}
\textcolor[rgb]{0,0,0}{The SE with perfect limited CSI in \eqref{eq:SVD-para-3}} can be expressed as
\begin{equation}\label{eq:R-PL}
  \begin{aligned}
{R_{{\rm{PL}}}} &= {\log _2}\det \left( {{\bf{I}}_{{N_{\rm R}}\times{N_{\rm R}}} + \frac{1}{{{\sigma ^2}}}{{\bf{W}}^H}{\bf{HF}}{{\bf{F}}^H}{{\bf{H}}^H}{\bf{W}}} \right)\\
& \buildrel {(a)} \over \approx {\log _2}\det \left( {\bf{I}}_{{N_{\rm R}}\times{N_{\rm R}}} + \frac{1}{{{\sigma ^2}}}{\bf{A}}_{{\rm{R,}} \bot }^H{{\bf{A}}_{{\rm{R,}} \bot }}{{\bf{\Xi }}_ \bot }{\bf{A}}_{{\rm{T,}} \bot }^H{{\bf{A}}_{{\rm{T,}} \bot }}\right.\\
&\quad\qquad\qquad\qquad\qquad\times\left.{\bf P}{\bf{A}}_{{\rm{T,}} \bot }^H{{\bf{A}}_{{\rm{T,}} \bot }}{\bf{\Xi }}_ \bot ^H{\bf{A}}_{{\rm{R,}} \bot }^H{{\bf{A}}_{{\rm{R,}} \bot }} \right)\\
&\buildrel {(b)} \over \approx \sum\limits_{k \in {{\mathcal K}_ \bot }} {{{\log }_2}\left( {1 + {{\left| {\xi _{k,{l_k},0}^\star } \right|}^2}{p_k}/{\sigma ^2}} \right)}.
\end{aligned}
\end{equation}
where $(a)$ and $(b)$ are due to ${\bf{H}}\approx {{\bf{A}}_{{\rm{R,}} \bot }}{{\bf{\Xi }}_ \bot }{\bf{A}}_{{\rm{T,}} \bot }^H$ and ${\bf{A}}_{{\rm{R,}} \bot }^H{{\bf{A}}_{{\rm{R,}} \bot }} \approx {\bf I}_{{ N_{{\rm R}} }\times{N_{{\rm R}}}}$, respectively.

In FDD systems, the limited CSI in \eqref{eq:SVD-para-3} has to be fed back to the Tx for enabling the design of the phase shifters and the Tx precoder. As mentioned in Section \ref{sec:3.1}, deploying the RISs at different DFT directions produces orthogonal LoS paths and reduces the feedback overhead. Note that ${\mathcal L}_{{\rm{T}},k}^\star  = \{ 0 \}$ for $k \in {\mathcal K}$ because the LoS paths are selected to construct ${{\bf{A}}_{{\rm{T,}} \bot }}$; $\{\Theta _{{\rm{T}},k,{0}}^{\rm{A}},k \in {{\mathcal K}_ \bot }\}$ in \eqref{eq:SVD-para-3} are available at the Tx without feedback by availing of the constant AoAs in the LoS path. Hence, the limited CSI that should be fed back to the Tx is expressed as
\begin{equation}\label{eq:SVD-para-2}
  \left\{ {\left( {| {\xi _{k,{l_k},{0}}^\star } |},{\Theta _{{k,{\rm{R}}},{l_k}}^{\rm{D}}} \right),k \in {{\mathcal K}_ \bot },{l_k} \in {\mathcal L}_{{k,{\rm{R}}}}^\star } \right\}.
\end{equation}
\textcolor[rgb]{0,0,0}{Compared with existing works that design the SVD transceiver with full CSI $\{ {{{\bf{H}}_{{k,{\rm{R}}}}},{{\bf{H}}_{{\rm{T}},k}}} \}_{k = 1}^K$ constructed by all channel parameters given by \eqref{eq:para_T} and \eqref{eq:para_R}, the feedback overhead in our proposed scheme is reduced from $3K+3{\sum\nolimits_{k=1}^{K}({{L_{{\rm{T}},k}} + {L_{{k,{\rm{R}}}}}})} $ to $2N_{\rm R}$. Considering that $K\ge N_{\rm R}$ and that the number of paths is usually large in the sub-6 GHz band, this overhead reduction is significant.}\par

Although the SVD transceiver is optimal with a given RIS design, we do not claim the optimality of our RIS design in terms of SE maximization. The optimal RIS design that maximizes the SE can be obtained by utilizing full CSI. However, this strategy is impractical because the feedback overhead and optimization complexity will be prohibitively high. The channel rank cannot be guaranteed as well. As a suboptimal solution, the proposed channel customization-based SVD transceiver reduces the feedback overhead and simplifies the system design (RIS design with statistical CSI and SVD transceiver without a matrix decomposition). The composite channel with full row rank can be customized to provide the spatial multiplexing.\par
Although the feedback overhead is greatly reduced, the limited CSI in \eqref{eq:SVD-para-2} should be further quantized at the Rx before feeding back to the Tx because of the limited feedback ability of the former. On this basis, we study channel feedback with limited quantization and its effects on the ergodic SE in the following section.

\section{Channel Feedback with Limited Quantization}\label{sec:4}
In this section, we first investigate the effect of limited CSI on the channel customization-based SVD transceiver design. We show that the water-filling algorithm can be approximately replaced by equal power allocation and the effect of singular values' feedback is averaged out by properly selecting the number of RIS elements. On this basis, the ergodic SE loss incurred by the limited quantization of \textcolor[rgb]{0,0,0}{directional parameters} is investigated. A closed-form approximation of the ergodic SE loss is derived, and a bit partitioning strategy is then developed for SE loss reduction.\par
\subsection{Effect of Singular Values}\label{sec:4.1}
Equal power allocation is asymptotically optimal in the high SNR regime \cite{MIMO_OFDM}. In multi-RIS assisted systems, carefully selecting the number of elements in each RIS to combat the path loss can degenerate the water-filling algorithm into equal power allocation even in the low SNR regime. We denote ${\xi _{0}^\star }$ as a constant and ${ {\xi _{k,{l_k},0}^{\rm r} } } = { {{\rho^{\rm r} _k}{\alpha _{{\rm{T}},k,0}}{\alpha _{{k,{\rm{R}}},{l_k}}}} }$ as the reference cascaded path gain in the Tx--RIS $k$--Rx channel, where ${\rho^{\rm r} _k} = \frac{\lambda }{{4\pi {d_{{\rm T},k}}}}\frac{\lambda }{{4\pi {d^{\rm r}_{{k,{\rm R}}}}}}$ with ${d^{\rm r}_{{k,{\rm R}}}}$ being the constant distance from RIS $k$ to the center of the blind coverage area. Given that ${\rho^{\rm r} _k}$ and ${\alpha _{{\rm{T}},k,0}}$ are constants, the average value of $|{ {\xi _{k,{l_k},0}^{\rm r} } }|^2$ is expressed as
\begin{equation}\label{eq:xi-r}
\begin{aligned}
  {\mathbb E}\left\{|{\xi _{k,{l_k},0}^{\rm r} }|^2\right\}&=\left({\rho^{\rm r} _k}{\alpha _{{\rm{T}},k,0}}\right)^2{\mathbb E}\left\{|{ {{\alpha _{{k,{\rm{R}}},{l_k}}}} }|^2\right\}\\
  &=\left({\rho^{\rm r} _k}{\alpha _{{\rm{T}},k,0}}\right)^2 {\frac{{{N_{\rm{R}}}{N_{{\rm{S,}}k}}}}{{{L_{{k,{\rm{R}}}}}}}}{\mathbb E}\left\{ |{\beta _{{k,{\rm{R}}},l_k}}|^2\right\}\\
  &\buildrel {(a)} \over =\left({\rho^{\rm r} _k}{\alpha _{{\rm{T}},k,0}}\right)^2 {\frac{{{N_{\rm{R}}}{N_{{\rm{S,}}k}}}}{{{L_{{k,{\rm{R}}}}}}}},
\end{aligned}
\end{equation}
where $(a)$ is due to ${\beta _{{k,{\rm{R}}},l_k}} \sim {\mathcal {CN}}(0,1)$. To obtain comparable cascaded path gains for all RISs, we let
\begin{equation}\label{eq:E-xi-r}
  {\mathbb E}\left\{|{\xi _{k,{l_k},0}^{\rm r} }|^2\right\}  \buildrel \Delta \over =  \xi _{0}^\star,\forall k.
\end{equation}
The number of elements for RIS $k$ by expanding ${\rho^{\rm r} _k}$ and ${\alpha _{{\rm{T}},k,0}}$ in ${\mathbb E}\{|{\xi _{k,{l_k},0}^{\rm r} }|^2\}$ is approximately given as
\begin{equation}\label{eq:N_s,k}
  {N_{{\rm{S,}}k}} = \left\lceil \frac{{16{\pi ^2}{d_{{\rm T},k}}d_{{k,{\rm R}}}^{\rm{r}}  }}{{{\lambda ^2}}}\sqrt {\frac{{\left( {{\kappa _{{\rm{T}},k}} + 1} \right){L_{{k,{\rm{R}}}}}{\xi _0^\star}}}{{{\kappa _{{\rm{T}},k}}{N_{\rm{T}}}{N_{\rm{R}}}}}}\right\rceil,
\end{equation}
where $\lceil\cdot\rceil$ is included because ${N_{{\rm{S,}}k}}$ is an integer. Once the communication system is finalized so that the environment and system parameters remain unchanged, the size of RISs can be flexibly determined by the constant ${\xi _0^\star}$. Now, $\{{| {\xi _{k,{l_k},{0}}^\star } |},k \in {{\mathcal K}_ \bot }\}$ in \eqref{eq:SVD-para-3} will be in the same order of magnitude by  combining \eqref{eq:N_s,k} and path pruning \eqref{eq:path-pruning} that omits weak paths. Thus, ${p_k} = \max ( {\mu  - {\sigma ^2}/{{| {\xi _{k,{l_k},0}^\star } |}^2},0} )$ can be approximately replaced by equal power allocation ${p_k} \approx E/|{{\mathcal K}_ \bot }|$ and the superiority of water-filling power allocation will become negligible. Although the equal power allocation underperforms the water-filling allocation in terms of SE, we will show in the numerical results that the performance deterioration is acceptable due to the RIS size constraint in \eqref{eq:N_s,k} that seeks to provide homogenous singular values.

\subsection{Effect of \textcolor[rgb]{0,0,0}{Directional Parameters}' Quantization}
Considering that singular values have insignificant effects, in the following, they are assumed to be known at the Tx to better characterize the effect of quantization for the following \textcolor[rgb]{0,0,0}{directional parameters}
\begin{equation}\label{eq:SVD-para-1}
  \left\{ {\left( {\Theta _{{k,{\rm{R}}},{l_k}}^{\rm{D}}} \right),k \in {{\mathcal K}_ \bot },{l_k} \in {\mathcal L}_{{k,{\rm{R}}}}^\star } \right\}.
\end{equation}
Considering that $| {{\mathcal L}_{{k,{\rm{R}}}}^\star } | = | {{\mathcal L}_{{\rm{T}},k}^\star } | = 1$, the subscript $l_{k}$ for the limited statistical CSI in \eqref{eq:SVD-para-1} can be omitted for the sake of presentation. Similarly, ${\xi _{k,{l_k},0}^\star }$ is simplified as ${\xi _{k}^\star }$. We concentrate on the feedback of the \textcolor[rgb]{0,0,0}{directional parameters} $\{ { {\Theta _{{k,{\rm{R}}}}^{\rm{D}}} } \}_{k = 1}^{{N_{\rm{R}}}}$, since the RISs set ${\mathcal K}_{\bot}$ containing the finite RISs indices can be easily and precisely fed back to the Tx.\par
With the perfect limited CSI $\{ {( {| {\xi _k^\star } |,\Theta _{{k,{\rm{R}}}}^{\rm{D}}} )} \}_{i = 1}^{{N_{\rm{R}}}}$, by performing the water-filling algorithm\footnote{The water-filling power allocation is used as a benchmark while its performance gain over equal power allocation is demonstrated in the numerical results.} to obtain the power allocation coefficients $\{ {p_k^\star } \}_{k = 1}^{{N_{\rm{R}}}}$ and applying Jensen's inequality on \eqref{eq:R-PL}, an upper bound on the ergodic SE can be derived as\footnote{In the following, the subscripts ``PL'' and ``QL'' are used to indicate the perfect and quantized limited CSI cases, respectively.}
\begin{equation}\label{eq:R_PL_u}
   {\mathbb E}\left\{R_{\rm PL}\right\}\le \sum\limits_{k = 1}^{{N_{\rm{R}}}} {{{\log }_2}\left( {1 + \frac{{\mathbb E}{\left\{ {{{\left| {\xi _k^\star } \right|}^2} }{p_k^\star } \right\}}}{\sigma ^2}} \right)}  \buildrel \Delta \over =  {R_{{\rm{PL,upper}}}}.
\end{equation}
When $\{ { {\Theta _{{k,{\rm{R}}}}^{\rm{D}}} } \}_{k = 1}^{{N_{\rm{R}}}}$ are quantized, the RIS design \eqref{eq:optimal-RIS} with quantized \textcolor[rgb]{0,0,0}{directional parameters} $\{ { {\tilde \Theta _{{k,{\rm{R}}}}^{\rm{D}}} } \}_{k = 1}^{{N_{\rm{R}}}}$ will produce the actual singular values as $\{ {{{| {{\xi _k}} |}}}= | {{\rho _k}{\alpha _{{k,{\rm{R}}}}}{\alpha _{{\rm{T}},k,0}}} {\bf{a}}_{{\rm{S,}}k}^H\left( {\Theta _{{k,{\rm{R}}}}^{\rm{D}} - {\tilde \Theta} _{{k,{\rm{R}}} }^{\rm{D}}} \right){{\bf{a}}_{{\rm{S}},k}}\left( 0 \right) |\}_{k = 1}^{{N_{\rm{R}}}}$. Hence, the upper bound ergodic SE with quantized limited CSI $\{ {( {| {\xi _k^\star } |,{\tilde \Theta} _{{k,{\rm{R}}}}^{\rm{D}}} )} \}_{i = 1}^{{N_{\rm{R}}}}$ can be expressed as
\begin{equation}\label{eq:R_QL_u}
   {R_{{\rm{QL,upper}}}} \buildrel \Delta \over = \sum\limits_{k = 1}^{{N_{\rm{R}}}} {{{\log }_2}\left( 1 + \frac{{\mathbb E}{\left\{ {{{\left| {\xi _k } \right|}^2} }{p_k^\star } \right\}}}{\sigma ^2} \right)}    .
\end{equation}
We define the ergodic SE loss as
\begin{equation}\label{eq:rate-loss}
\begin{aligned}
  \Delta R &= {R_{{\rm{PL,upper}}}} - {R_{{\rm{QL,upper}}}}
  = \sum\limits_{k = 1}^{{N_{\rm{R}}}} {{{\log }_2}\frac{{{\sigma ^2} +{\mathbb E}\left\{ {{{\left| {\xi _k^\star } \right|}^2{p_k^\star }}} \right\}}}{{{\sigma ^2} + {\mathbb E}\left\{ {{{\left| {{\xi _k}} \right|}^2{p_k^\star }}} \right\}}}}   .
\end{aligned}
\end{equation}
Assume that the transmit power at the Tx is sufficient to guarantee the transmission of $N_{\rm R}$ data streams, that is, ${p_k^\star }>0$, $\forall k\in \{1,2,\ldots,N_{\rm R}\}$. In this case, $p_k^\star  = {\mu ^\star } - {{{\sigma ^2}}}/{{{{| {\xi _k^\star } |}^2}}}$, where the water level that satisfies $\sum\nolimits_{k = 1}^{{N_{\rm{R}}}} {{p_k}}  = E$ can be obtained as
\begin{equation}\label{eq:mu-tag}
  {\mu ^\star } = \frac{E}{{{N_{\rm{R}}}}} + \sum\nolimits_{i = 1}^{{N_{\rm{R}}}} {\frac{{{\sigma ^2}}}{{{N_{\rm{R}}}{{\left| {\xi _i^\star } \right|}^2}}}} .
\end{equation}
Substituting $p_k^\star$ into \eqref{eq:rate-loss} yields
\begin{equation}\label{eq:rate-loss-1}
  \begin{aligned}
  	&\Delta R =\\
  	&  \sum\limits_{k = 1}^{{N_{\rm{R}}}} {{{\log }_2}\left( {\frac{{1 + {\mathbb E}\left\{ {{C_k}} \right\}}}{{1 + {\mathbb E}{\left\{ {{C_k}{{\left| {{\bf{a}}_{{\rm{S}},k}^H\left( {\Theta _{{k,{\rm{R}}}}^{\rm{D}} - \tilde \Theta _{{k,{\rm{R}}}}^{\rm{D}}} \right){{\bf{a}}_{{\rm{S}},k}}\left( 0 \right)} \right|}^2}} \right\}}}}} \right)},
  \end{aligned}
\end{equation}
where
\begin{equation}\label{eq:C-k}
  {C_k} = {\left| {\xi _k^\star } \right|^2}\left( {\frac{E}{{{N_{\rm{R}}}{\sigma ^2}}} + \sum\nolimits_{i = 1}^{{N_{\rm{R}}}} {\frac{1}{{{N_{\rm{R}}}{{\left| {\xi _i^\star } \right|}^2}}}}  - \frac{1}{{{{\left| {\xi _k^\star } \right|}^2}}}} \right),
\end{equation}
is a function of Gaussian variables $\{\beta_{{k,{\rm R}}}\}^{N_{\rm R}}_{k=1}$. Utilizing the independence between the parameters $\{\beta_{{k,{\rm R}}}\}^{N_{\rm R}}_{k=1}$ and $\{\Theta^{\rm D}_{{k,{\rm R}}}\}^{N_{\rm R}}_{k=1}$, and the equality ${{{| {{\bf{a}}_{{\rm{S}},k}^H( {\Theta _{{k,{\rm{R}}}}^{\rm{D}} - \tilde \Theta _{{k,{\rm{R}}}}^{\rm{D}}} ){{\bf{a}}_{{\rm{S}},k}}( 0 )} |}^2}} = {\frac{{{{\sin }^2}( {{N_{{\rm{S}},k}}( {\tilde \Theta _{{k,{\rm{R}}}}^{\rm{D}} - \Theta _{{k,{\rm{R}}}}^{\rm{D}}} )/2} )}}{{N_{{\rm{S}},k}^2{{\sin }^2}( {( {\tilde \Theta _{{k,{\rm{R}}}}^{\rm{D}} - \Theta _{{k,{\rm{R}}}}^{\rm{D}}} )/2} )}}}$ \cite{bit_par}, \eqref{eq:rate-loss-1} can be further rewritten as
\begin{equation}\label{eq:rate-loss-2}
  \Delta R  = \sum\limits_{k = 1}^{{N_{\rm{R}}}} {{{\log }_2}\left( {\frac{{1 + {\mathbb E}\left\{ {{C_k}} \right\}}}{{1 + {\mathbb E}{\left\{ {{C_k}} \right\}}{\mathbb E}{\left\{ {\frac{{{{\sin }^2}\left( {{N_{{\rm{S}},k}}\left( {\tilde \Theta _{{k,{\rm{R}}}}^{\rm{D}} - \Theta _{{k,{\rm{R}}}}^{\rm{D}}} \right)/2} \right)}}{{N_{{\rm{S}},k}^2{{\sin }^2}\left( {\left( {\tilde \Theta _{{k,{\rm{R}}}}^{\rm{D}} - \Theta _{{k,{\rm{R}}}}^{\rm{D}}} \right)/2} \right)}}} \right\}}}}} \right)}.
\end{equation}
The expectation of $C_k$, denoted as ${\bar C}_k={\mathbb E}\left\{ {{C_k}} \right\}$, is a finite constant that indicates the average link quality of the $k$-th data streams and can be approximated as (Appendix \ref{App:A})
\begin{equation}\label{eq:E-C}
  {\bar C}_k  \approx {\frac{E}{{{N_{\rm{R}}}{\sigma ^2}}} } {\left( {\frac{{d_{2,k}^{\rm{r}}\xi _0^\star }}{{{d_{2,k}}}}} \right)^2}.
\end{equation}
The SE loss is determined by the limited quantization errors $\{{\tilde \Theta _{{k,{\rm{R}}}}^{\rm{D}} - \Theta _{{k,{\rm{R}}}}^{\rm{D}}}\}^{N_{\rm R}}_{k=1} $. Without loss of generality, the number of feedback bits, which affect the quantization error, is denoted by $B$.

\subsection{Approximation of the SE Loss}
To characterize the explicit effect of feedback bits, a closed-form expression for the SE loss is derived in this subsection by approximating the expectation of trigonometric functions.\par

\begin{figure}[!t]
\centering
\includegraphics[width=0.5\textwidth]{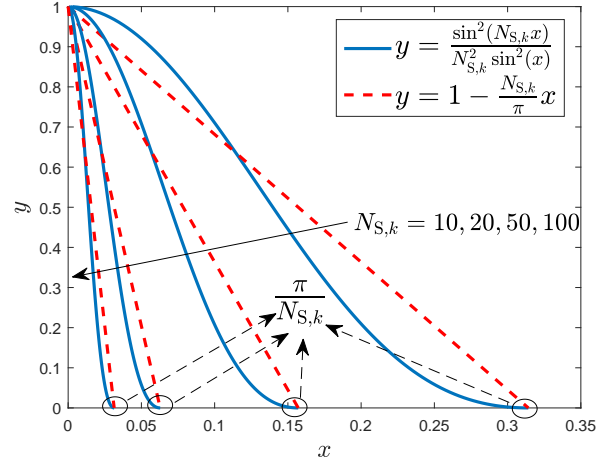}
\caption{Comparison of the trigonometric and linear functions.}
\label{Fig.func-appx} 
\end{figure}

When the size of the RISs increases, a higher spatial resolution will be achieved. To avoid the beam deviation when designing the RIS with quantized CSI, the number of feedback bits is proportional to the number of elements per RIS. As shown in \eqref{eq:rate-loss-2}, a larger value for ${\frac{{{{\sin }^2}( {{N_{{\rm{S}},k}}( {\tilde \Theta _{{k,{\rm{R}}}}^{\rm{D}} - \Theta _{{k,{\rm{R}}}}^{\rm{D}}} )/2} )}}{{N_{{\rm{S}},k}^2{{\sin }^2}( {( {\tilde \Theta _{{k,{\rm{R}}}}^{\rm{D}} - \Theta _{{k,{\rm{R}}}}^{\rm{D}}} )/2} )}}}$ shrinks the SE loss. The function ${\frac{{{{\sin }^2}( {{N_{{\rm{S}},k}}( {\tilde \Theta _{{k,{\rm{R}}}}^{\rm{D}} - \Theta _{{k,{\rm{R}}}}^{\rm{D}}} )/2} )}}{{N_{{\rm{S}},k}^2{{\sin }^2}( {( {\tilde \Theta _{{k,{\rm{R}}}}^{\rm{D}} - \Theta _{{k,{\rm{R}}}}^{\rm{D}}} )/2} )}}}$  monotonically decreases when $\frac{\tilde \Theta _{{k,{\rm{R}}}}^{\rm{D}} - \Theta _{{k,{\rm{R}}}}^{\rm{D}}}{2} \in [0,\frac{\pi}{N_{{\rm S},{k}}}]$. Starting from this property, we assume that a minimal number of feedback bits, ${b_{\min ,k}}$,  is required so that the maximum quantization error will not exceed the first null $\frac{\pi}{N_{{\rm S},{k}}}$, that is,
\begin{equation}\label{eq:first-null}
  \frac{{\left| {\tilde \Theta _{{k,{\rm{R}}}}^{\rm{D}} - \Theta _{{k,{\rm{R}}}}^{\rm{D}}} \right|_{\max}}}{2} \le \frac{\pi }{{{N_{{\rm{S}},k}}}}.
\end{equation}
With $\Theta _{{k,{\rm{R}}}}^{\rm{D}} = \pi\cos \theta _{{k,{\rm{R}}}}^{\rm{D}} \in [ { - \pi,\pi} ]$, ${b_{\min ,k}}$-bits quantization for $\Theta _{{k,{\rm{R}}}}^{\rm{D}}$ generates uniform grids of size ${2\pi}/{2^{b_{\min,k}}}$. Thus, the maximum distance between $\tilde\Theta _{{k,{\rm{R}}}}^{\rm{D}}$ and $\Theta _{{k,{\rm{R}}}}^{\rm{D}}$ will be
\begin{equation}\label{eq:max-error}
  {{{{\left| {\tilde \Theta _{{k,{\rm{R}}}}^{\rm{D}} - \Theta _{{k,{\rm{R}}}}^{\rm{D}}} \right|}_{\max }}}} = \frac{{2\pi }}{{ {2^{{b_{\min ,k}}+1}}}}.
\end{equation}
Combining \eqref{eq:first-null} and \eqref{eq:max-error}, the minimum number of feedback bits is then given as
\begin{equation}\label{eq:b-min}
  {b_{\min ,k}} =\left\lceil {{{\log }_2}\frac{{{N_{{\rm{S}},k}}}}{2 }} \right\rceil .
\end{equation}
With the minimum feedback bits, we have $x \buildrel \Delta \over = \frac{{\left| {\tilde \Theta _{{k,{\rm{R}}}}^{\rm{D}} - \Theta _{{k,{\rm{R}}}}^{\rm{D}}} \right|}}{2}\in [0,\frac{\pi}{N_{{\rm S},k}}]$. In this interval, the trigonometric function $\frac{\sin ^2(N_{{\rm S},k} x)}{{N_{{\rm S},k}^2}\sin^2(x)}$ can be approximated by the linear function $1-\frac{N_{{\rm S},k}}{\pi}x$ when $N_{{\rm S},k}$ increases, as shown in Fig. \ref{Fig.func-appx}. Utilizing this approximation, the expectation of the trigonometric function inside $\Delta R$ can be expressed as
\begin{equation}\label{eq:E-approx}
	\begin{aligned}
		&{\mathbb E}{\left\{ {\frac{{{{\sin }^2}\left( {{N_{{\rm{S}},k}}\left( {\tilde \Theta _{{k,{\rm{R}}}}^{\rm{D}} - \Theta _{{k,{\rm{R}}}}^{\rm{D}}} \right)/2} \right)}}{{N_{{\rm{S}},k}^2{{\sin }^2}\left( {\left( {\tilde \Theta _{{k,{\rm{R}}}}^{\rm{D}} - \Theta _{{k,{\rm{R}}}}^{\rm{D}}} \right)/2} \right)}}} \right\}}\\
		\approx &{\mathbb E}\left\{1 - \frac{{{N_{{\rm{S}},k}}}}{\pi }\frac{{\left| {\tilde \Theta _{{k,{\rm{R}}}}^{\rm{D}} - \Theta _{{k,{\rm{R}}}}^{\rm{D}}} \right|}}{2}\right\}.
	\end{aligned}
\end{equation}
\par
Given that $N_{\rm R}$ \textcolor[rgb]{0,0,0}{directional parameters} should be fed back, we allocate ${b_k} = {b_{\min ,k}} + {x_k}$  (${x_k} \in {{\mathbb Z}^ + }$) bits for quantizing $\Theta _{{k,{\rm{R}}}}^{\rm{D}}$ with $\sum\nolimits^{N_{\rm R}}_{k=1}b_k=B$. The quantization error then satisfies
\begin{equation}\label{eq:error-x}
  \frac{{\tilde \Theta _{{k,{\rm{R}}}}^{\rm{D}} - \Theta _{{k,{\rm{R}}}}^{\rm{D}}}}{2} \in \left[ { - \frac{{{2^{ - {x_k}}}\pi }}{{{N_{{\rm{S}},k}}}},\frac{{{2^{ - {x_k}}}\pi }}{{{N_{{\rm{S}},k}}}}} \right].
\end{equation}
In the quantization error interval, the expectation of of the linear function can be easily calculated as
\begin{equation}\label{eq:E-linear}
  {\mathbb E}\left\{1 - \frac{{{N_{{\rm{S}},k}}}}{\pi }\frac{{\left| {\tilde \Theta _{{k,{\rm{R}}}}^{\rm{D}} - \Theta _{{k,{\rm{R}}}}^{\rm{D}}} \right|}}{2}\right\}={1 - {2^{ - 1 - {x_k}}}}.
\end{equation}
Thus, the SE loss can be approximated as
\begin{equation}\label{eq:rate-loss-close}
  \Delta R  \approx \sum\limits_{k = 1}^{{N_{\rm{R}}}} {{{\log }_2}\left( {\frac{{1 +  {{{\bar C}_k}} }}{{1 + {{\bar C}_k}\left({1 - {2^{ - 1 - {x_k}}}}\right)}}} \right)}.
\end{equation}
Considering that ${1 - {2^{ - 1 - {x_k}}}}<1$, the SE loss can be upper bounded as
\begin{equation}\label{eq:rate-loss-upp}
\begin{aligned}
  \Delta R  &\approx \sum\limits_{k = 1}^{{N_{\rm{R}}}} {{{\log }_2}\left( {\frac{{1 +  {{{\bar C}_k}} }}{{1 + {{\bar C}_k}\left({1 - {2^{ - 1 - {x_k}}}}\right)}}} \right)}\\
  &<\sum\limits_{k = 1}^{{N_{\rm{R}}}} {{{\log }_2}\left( {\frac{{1    }}{{ {1 - {2^{ - 1 - {x_k}}}}}}} \right)}.
\end{aligned}
\end{equation}
Therefore, when the minimum number of feedback bits is allocated, that is, $x_k=0$, each data stream experiences $1$ bps/Hz SE loss at most. Increasing the feedback bits reduces the SE loss. For fixed-size RISs, $x_k \to \infty$ holds, and $\Delta R\to 0$ can be achieved when the number of feedback bits goes to infinity, which is consistent with the ideal case that perfect limited CSI is available. However, the total number of feedback bits is limited in practice. Given a fixed number of feedback bits that satisfy the minimum requirement, that is, $B>B_{\min}={\sum\nolimits^{N_{\rm R}}_{k=1}b_{\min,k}}$, a bit partitioning strategy for the extra feedback bits ($B-B_{\min}$) should be developed to reduce the SE loss.

\subsection{Bit Partitioning}
To minimize the SE loss, we first rewrite \eqref{eq:rate-loss-close} as
\begin{equation}\label{eq:rate-loss-rw}
  \Delta R  \approx \sum\limits_{k = 1}^{{N_{\rm{R}}}} {{{\log }_2}\left( {{{1 +  {{{\bar C}_k}} }}} \right)} - \sum\limits_{k = 1}^{{N_{\rm{R}}}} {{{\log }_2}\left( {{{1 + {{\bar C}_k}\left({1 - {2^{ - 1 - {x_k}}}}\right)}}} \right)}.
\end{equation}
Given that $\{{\bar C}_k\}^{N_{\rm R}}_{k=1}$ are constants and independent of the quantization for \textcolor[rgb]{0,0,0}{directional parameters}, the bit partitioning that minimizes the SE loss can be formulated as
\begin{equation}\label{eq:opt-max}
  \begin{aligned}
\mathop {\max }\limits_{\left\{ {{x_k}} \right\}_{k = 1}^{{N_{\rm{R}}}}} &\sum\limits_{k = 1}^{{N_{\rm{R}}}} {{{\log }_2}\left( {{{1 + {{\bar C}_k}\left({1 - {2^{ - 1 - {x_k}}}}\right)}}} \right)} \\
{\rm{s.t.}}\quad&{x_k} \in {{\mathbb Z}^ + }\\
&\sum\limits_{k = 1}^{{N_{\rm{R}}}} {{x_k}}  = {B} - {B_{{\rm{min}}}}.
\end{aligned}
\end{equation}
The maximization process is a combinational problem, whose optimal solution through exhaustive search has a complexity of $N_{\rm{R}}^{{B} - {B_{{\rm{min}}}}}$. To reduce the search complexity, we resort to greedy search. Given that the objective function in \eqref{eq:opt-max} is the sum of increasing functions of $x_k$, every additional bit will increase its value. \textcolor[rgb]{0,0,0}{Therefore, the original problem can be equivalently decoupled into ${{B} - {B_{{\rm{min}}}}}$ sequential $1$-bit assignment sub-problems and solved by greedy search.} For the $i$-th sub-problem, $i-1$ extra bits are allocated, that is, $\sum\nolimits_{k = 1}^{{N_{\rm{R}}}} {{x_k}}  = i-1$. The $1$-bit assignment in the $i$-th sub-problem can be expressed as
\begin{equation}\label{eq:opt-1bit}
  \begin{aligned}
\mathop {\max }\limits_{\left\{ {{n_k}} \right\}_{k = 1}^{{N_{\rm{R}}}}} &\sum\limits_{k = 1}^{{N_{\rm{R}}}} {{{\log }_2}\left( {{{1 + {{\bar C}_k}\left({1 - {2^{ - 1 - ({x_k}+n_k)}}}\right)}}} \right)} \\
{\rm{s.t.}}\quad&{n_k} \in {{\mathbb Z}^ + }\\
&\sum\limits_{k = 1}^{{N_{\rm{R}}}} {{n_k}}  = 1.
\end{aligned}
\end{equation}
For the $k$-th data stream, one additional bit on $x_k$, that is, $n_k=1$, can obtain a SE increase, which can be expressed as
\begin{equation}\label{eq:promo-1bit}
\begin{aligned}
  {f_k}\left( {{x_k}} \right) = &{\log _2}\left({{1 + {{\bar C}_k}\left( {1 - {2^{ - 1 - \left( {{x_k} + 1} \right)}}} \right)}}\right) \\
  &- {\log _2}\left({{1 + {{\bar C}_k}\left( {1 - {2^{ - 1 - {x_k}}}} \right)}}\right)\\
   = &{\log _2}\left( {1 + \frac{1}{{{2^{2 + {x_k}}}\left( {1/{{\bar C}_k} + 1} \right) - 2}}} \right).
\end{aligned}
\end{equation}
With the definition of SE increase, we explore insights into reducing the SE loss with bit partitioning.\par
Note that ${f_k}\left( {{x_k}} \right) \ge {f_m}\left( {{x_m}} \right)$ should hold for any $ m\in \left\{1,2,\ldots,N_{\rm R}\right\}\backslash\left\{k\right\}$ to obtain the largest SE increase for the $k$-th data stream with one more feedback bit. Expanding ${f_k}\left( {{x_k}} \right) \ge {f_m}\left( {{x_m}} \right)$ for $\forall m\in \left\{1,2,\ldots,N_{\rm R}\right\}\backslash\left\{k\right\}$, the current number of extra bits $x_k$ should satisfy
\begin{equation}\label{eq:xk-xm}
  {x_k}  + {\log _2}\left({{\frac{1}{{\bar C}_k} + 1}}\right) \leq {x_m}+ {\log _2}\left({{\frac{1}{{\bar C}_m} + 1}}\right).
\end{equation}
The condition \eqref{eq:xk-xm} provides a general criterion to address the sub-problem in \eqref{eq:opt-1bit}. \textcolor[rgb]{0,0,0}{With $N_{\rm R}-1$ comparisons in \eqref{eq:xk-xm}, obtaining the optimal bit partitioning by greedy search has a search complexity of $(N_{\rm R}-1)(B-B_{\rm min})$, which is much less than $N_{\rm R}^{B-B_{\rm min}}$ entailed by exhaustive search.} When the extra feedback bits $\{x_k\}^{N_{\rm R}}_{k=1}$ are the same, \eqref{eq:xk-xm} reduces to
\begin{equation}\label{eq:ck-cm}
  {{\bar C}_k} > {{\bar C}_m},\forall m\in \left\{1,2,\ldots,N_{\rm R}\right\}\backslash\left\{k\right\},
\end{equation}
indicating that the $k$-th data stream with the largest link quality has priority in the one extra feedback bit.\par
Also ${f_k}\left( {{x_k}} \right)$ is a decreasing function of $x_k$ and is upper bounded as
\begin{equation}\label{eq:f0}
  {f_k}\left( {{x_k}} \right) < {\log _2}\left( {1 + \frac{1}{{{2^{2 + x_k }} - 2}}} \right) = {f_{k,{\rm upper}}}\left( {{x_k}} \right).
\end{equation}
This property indicates that the SE increase provided by one more bit decreases with the increase in feedback bits and the largest SE increase is upper bounded by ${f_{k,{\rm upper}}}( {{0}} )={\log _2}(\frac{3}{2})$ bps/Hz. Considering that the upper bound of ${f_k}\left( {{x_k}} \right)$ can be approached by increasing ${\bar C}_k$, \eqref{eq:f0} reveals that when the extra feedback bits $\{x_k\}^{N_{\rm R}}_{k=1}$ are the same and $\{{\bar C}_k\}^{N_{\rm R}}_{k=1}$ are sufficiently large (which can be achieved by increasing the transmit power $E$ or the constant $\xi _0^\star$ that determines the number of RIS elements), the SE increase is equal for each data stream. In this case, the optimal extra bit partitioning for the original problem \eqref{eq:opt-max} is equal allocation, that is, ${x_k} = \frac{{{B} - {B_{\min}}}}{{{N_{\rm{R}}}}}$. When $\frac{{{B} - {B_{\min}}}}{{{N_{\rm{R}}}}}$ is not an integer, ${x_k} =\lfloor \frac{{{B} - {B_{\min}}}}{{{N_{\rm{R}}}}}\rfloor$ and the rest $(B - {B_{{\rm{min}}}} - \sum\nolimits_{k = 1}^{{N_{\rm{R}}}} {{x_k}} )$ can be randomly allocated to $(B - {B_{{\rm{min}}}} - \sum\nolimits_{k = 1}^{{N_{\rm{R}}}} {{x_k}} )$ data streams.

\section{Numerical Results}\label{sec:5}
\begin{figure}[!t]
\centering
\includegraphics[width=0.5\textwidth]{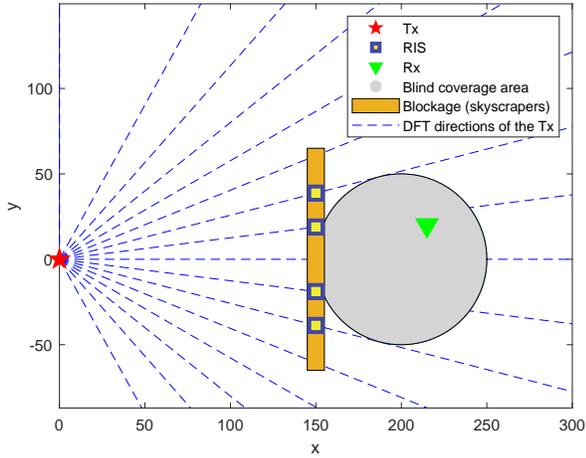}
\caption{Top view of the multi-RIS-assisted system deployment.}
\label{Fig.sys-set} 
\end{figure}

In this section, we present the numerical results to illustrate the effectiveness of the proposed channel customization in reducing the feedback overhead of multiple RIS-assisted sub-6 GHz FDD MIMO systems. The SE loss introduced by the limited feedback ability of the Rx is investigated. The system carrier frequency is set as $f_{\rm c}=3.5$ GHz. The numbers of antennas at the Tx and Rx are $N_{\rm{T}}=16$ and $N_{\rm R}=4$, respectively. As shown in Fig. \ref{Fig.sys-set}, we assume that the Tx is fixed at the original Cartesian coordinates and the Rx is randomly distributed in a blind coverage area denoted by a circle, whose center and radius are $[200,0]^T$ and $50$, respectively. A series of DFT directions for the Tx antenna array are represented by rays emitted from $[0,0]^T$. Four RISs are deployed between the Tx and Rx to establish the transmission link. All RISs are mounted on the DFT directions with $x$-coordinate being $150$ to create orthogonal LoS paths for the Tx--RISs channel and to reduce the feedback overhead, as discussed in Section \ref{sec:3}. Unless otherwise specified, the number of elements for each RIS is determined by \eqref{eq:N_s,k} with $\xi_0^\star=2.27\times 10^{-11}$. In other words, \textcolor[rgb]{0,0,0}{$N_{{\rm S},{1}}=N_{{\rm S},{4}}=416$ and $N_{{\rm S},{2}}=N_{{\rm S},{3}}=338$} will be set in accordance with the considered system deployment. Considering that the Tx and RISs are typically installed at high-rise buildings, the Rician factors are set to ${\kappa}_{{\rm T},k}=10$ dB, $\forall k$. The number of NLoS paths for the Tx--RISs and for the RISs--Rx channels are $L_{{\rm T},k}=2$ and $L_{{k,{\rm R}}}=10$, respectively. The AoA and AoD of NLoS paths are determined by the random scatterers between the Tx/Rx--RIS channel. For the path pruning, we set $\tau =10$ dB. The noise power is $-100$ dBm.

\subsection{Feedback Overhead Reduction with Channel Customization}
\begin{figure}[!t]
\centering
\subfigure[Before channel customization]{
\label{Fig.DPCC-be} 
\includegraphics[width=0.5\textwidth]{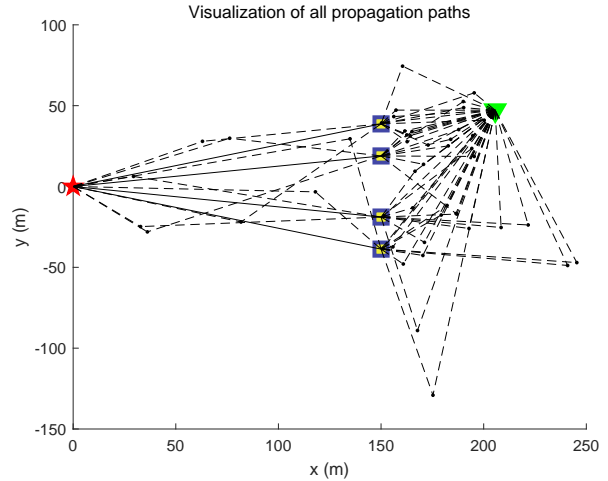}}
\subfigure[After channel customization]{
\label{Fig.DPCC-af} 
\includegraphics[width=0.5\textwidth]{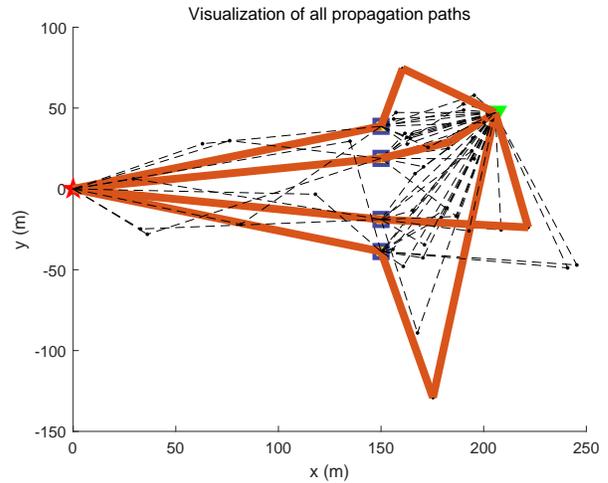}}
\caption{Propagation paths between the Tx and the Rx in a channel realization.}
\label{Fig.DPCC} 
\end{figure}

\begin{figure}[!t]
\centering
\includegraphics[width=0.5\textwidth]{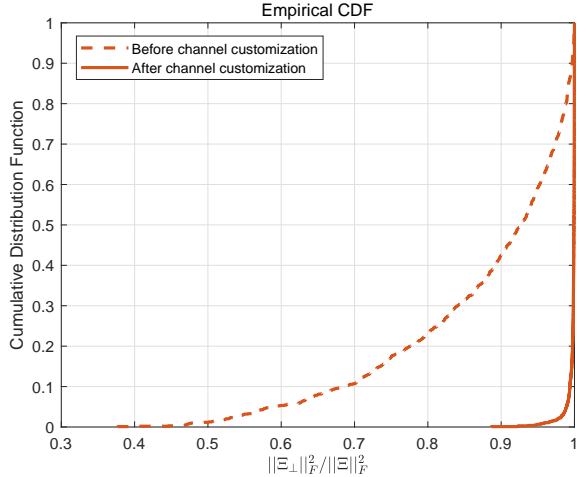}
\caption{CDF of the ratio between the power of the orthogonal paths over the power of the overall paths in $1000$ channel realizations.}
\label{Fig.path-energy} 
\end{figure}

Figure \ref{Fig.DPCC} visualizes all cascaded propagation paths between the Tx and Rx before and after RIS design in a channel realization to intuitively present the concept of channel customization that reduces feedback overhead. Considering that $3$ paths between the Tx--RIS $k$ channel and $10$ paths between the RIS $k$--Rx channel are found, $30$ cascaded paths can be built for each Tx--RIS $k$--Rx channel. Figure \ref{Fig.DPCC-be} shows that the raw composite channel without RIS design has $120$ meshed and weak paths, which entails a huge overhead for the feedback of channel parameters. Four orthogonal paths are first selected to reduce this overhead, as shown in Fig. \ref{Fig.DPCC-af}. We configure the phase shifters to enhance the corresponding path gains so that the composite channel can be dominated by the four orthogonal paths. Figure \ref{Fig.path-energy} presents the CDF of the ratio between the power of the orthogonal paths over the power of the overall paths in $1000$ channel realizations. We can observe that before channel customization, four paths with considerable path power can be picked through the path pruning and selection. However, these paths are inadequate to reconstruct the composite channel. With the proposed RIS design, the path power ratio is rapidly converging to approximately $1$ in all channel realizations, indicating that the composite channel can be well approximated by the orthogonal components that are made up of four dominant paths. The overhead is reduced by feeding back the parameters of the dominant paths only.

\subsection{SE Loss Reduction with Limited Feedback Bits}

\begin{figure}[!t]
\centering
\includegraphics[width=0.5\textwidth]{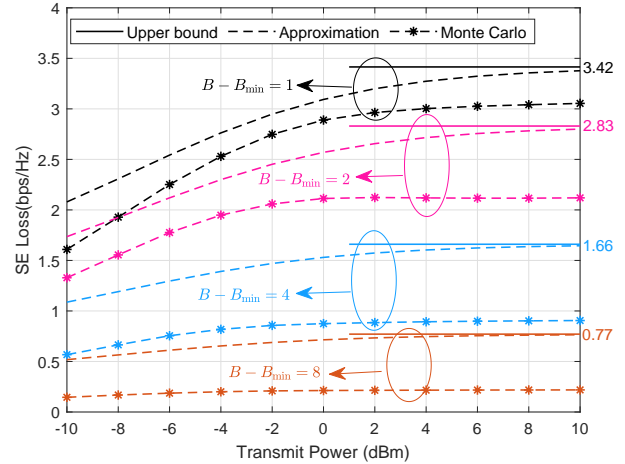}
\caption{Theoretical analysis and Monte Carlo result of the SE loss as a function of the transmit power for an increasing number of extra feedback bits $B-B_{\min}$.}
\label{Fig.rate-loss} 
\end{figure}

In this subsection, we assess the SE loss incurred by the limited feedback ability of the Rx. The minimum number of feedback bits $B_{\min}=\sum\nolimits ^{N_{\rm R}}_{k}{b_{{\min},k}}$ is determined by \eqref{eq:b-min}, and the total feedback bits $B$ can be embodied by
the extra bits $B-B_{\min}$. These extra bits are partitioned in accordance with \eqref{eq:xk-xm} to reduce the SE loss. Figure \ref{Fig.rate-loss} compares the upper bound in \eqref{eq:rate-loss-upp}, approximation in \eqref{eq:rate-loss-close}, and Monte Carlo result of the SE loss versus the transmit power when the extra feedback bits are increasing from $1$ to $8$. A performance gap is observed between the approximation and the Monte Carlo result. This condition is because the SE loss approximation is defined by the upper bound of the ergodic SE, and the Monte Carlo result is calculated by averaging the difference in the instantaneous SE in \eqref{eq:R} with the perfect and quantized CSI. Despite this performance gap, the approximation has the same tendency as the Monte Carlo result, which is helpful for reducing the SE loss. Figure \ref{Fig.rate-loss} shows that when the transmit power increases, the SE loss approximation will approach its upper bound, which coincides with our analysis in \eqref{eq:f0}. The SE loss can be continuously reduced by providing more extra feedback bits. Equation \eqref{eq:f0} reveals that in the high transmit power regime, the bit partitioning for extra bits will be simplified to equal allocation. Thus, Fig. \ref{Fig.rate-loss} shows that when the transmit power is sufficiently large, the SE loss approximations are close to the upper bounds $3+{\log_2}(\frac{4}{3}) \approx 3.42$, $2+2{\log_2}(\frac{4}{3})\approx 2.83$, $4{\log_2}(\frac{4}{3})\approx 1.66$, and $4{\log_2}(\frac{8}{7}) \approx 0.77$ when the extra bits are set to $1$, $2$, $4$, and $8$, respectively.\par

In the following, the legends \textbf{P-CSI}, \textbf{PL-CSI}, \textbf{QL-CSI}, \textbf{PLS-CSI}, and \textbf{QLS-CSI} refer to the scenarios where the Perfect CSI, Perfect Limited CSI ($\{ {( {| {\xi _k^\star } |,\Theta _{{k,{\rm{R}}}}^{\rm{D}}} )} \}_{i = 1}^{{N_{\rm{R}}}}$), Quantized Limited CSI ($\{ {( {| {\xi _k^\star } |,{\tilde \Theta} _{{k,{\rm{R}}}}^{\rm{D}}} )} \}_{i = 1}^{{N_{\rm{R}}}}$), Perfect Limited Statistical CSI ($\{ {( {\Theta _{{k,{\rm{R}}}}^{\rm{D}}} )} \}_{i = 1}^{{N_{\rm{R}}}}$), and Quantized Limited Statistical CSI ($\{ {( {{\tilde \Theta} _{{k,{\rm{R}}}}^{\rm{D}}} )} \}_{i = 1}^{{N_{\rm{R}}}}$) are available at the Tx, respectively.\par

\begin{figure}[!t]
	\centering
	\includegraphics[width=0.5\textwidth]{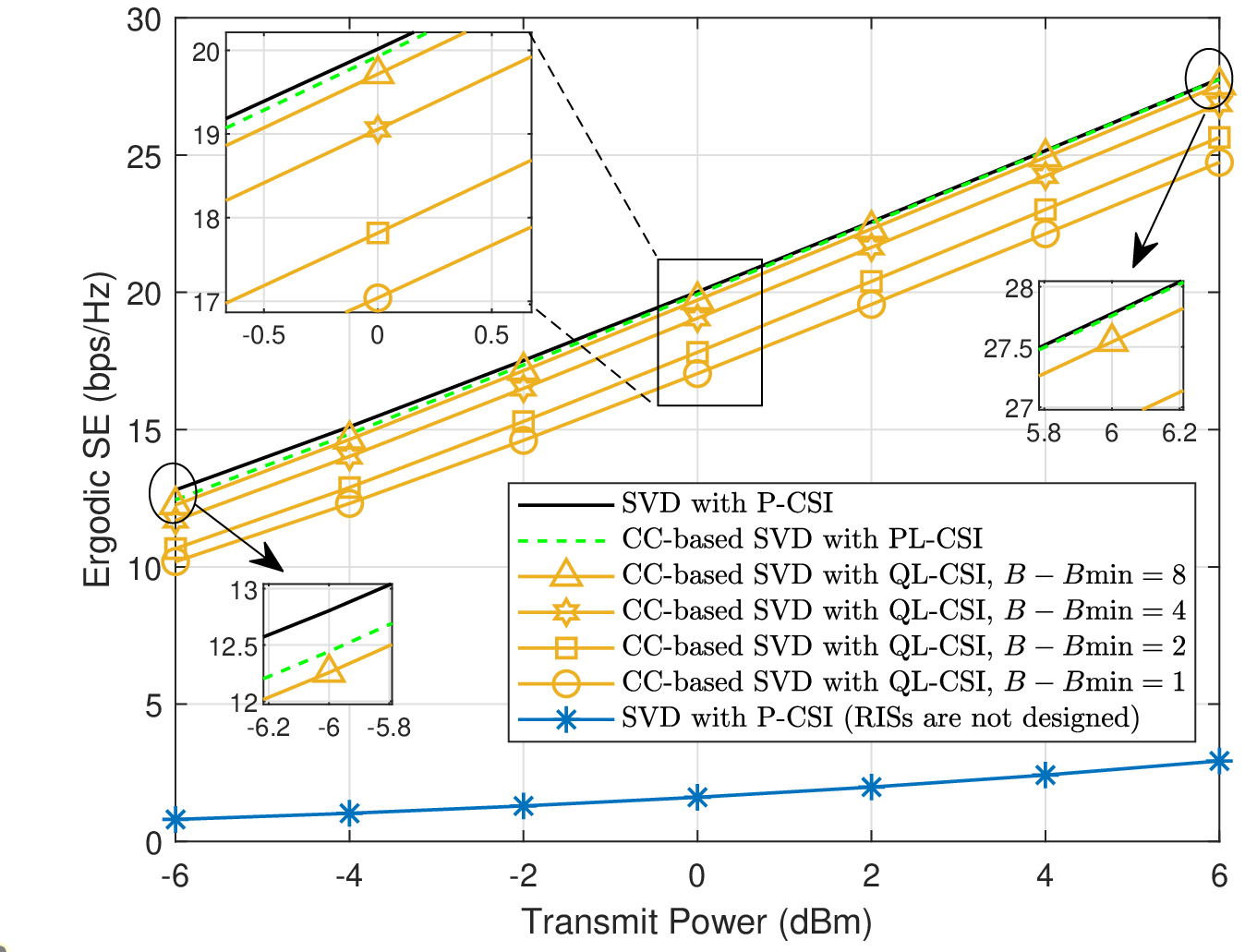}
	\caption{Ergodic SE versus transmit power with different feedback abilities.}
	\label{Fig.RVE} 
\end{figure}

Figure \ref{Fig.RVE} compares the ergodic SE, which is the average of \eqref{eq:R} over the channel realizations, of different SVD transceiver designs. The RISs are designed with different CSI levels to customize the composite channel. The SVD transceiver and channel customization-based (CC-based) SVD transceiver are designed in accordance with \eqref{eq:optimal_F_W} and \eqref{eq:CC-SVD}, respectively. From Fig. \ref{Fig.RVE}, the ergodic SE of the CC-based SVD with PL-CSI is inferior to that of the SVD with P-CSI in the low transmit power regime. However, this ergodic SE gap is squeezed, and the performance of the two transceivers becomes gradually equivalent with the increase in the transmit power. Considering that the CC-based transceiver requires only limited CSI and does not rely on channel reconstruction and matrix decomposition, this result is extremely promising because feedback overhead and computational complexity are significantly reduced. When the feedback bits are limited, quantization error appears in the \textcolor[rgb]{0,0,0}{directional parameters} that are fed back from the Rx and then used to design the phase shifters of RISs. The quantization error decreases the singular values of the customized channel. Therefore, the ergodic SE deteriorates. Given the minimum feedback bits, we can see that eight additional bits are sufficient to achieve a comparable performance compared with the infinite quantization case. When the RISs are not designed, the ergodic SE reduces intensively, as shown in Fig. \ref{Fig.RVE}. This huge performance gap emphasizes the importance of the proposed channel customization that provides an essential tool for transceiver design.\par

\begin{figure}[!t]
\centering
\includegraphics[width=0.5\textwidth]{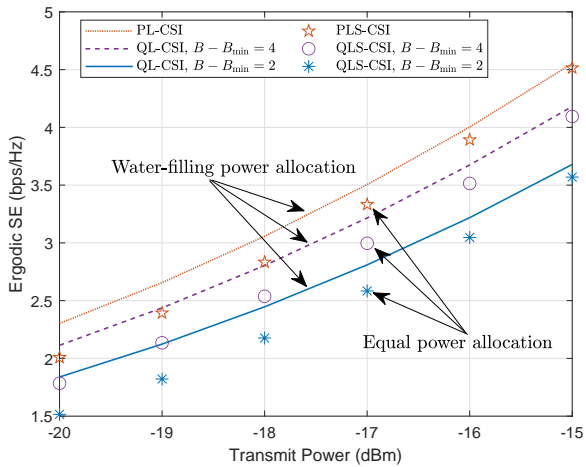}
\caption{Comparison of the ergodic SE with water-filling and equal power allocation.}
\label{Fig.water-equal} 
\end{figure}
The ergodic SEs under water-filling and equal power allocation are compared in Fig. \ref{Fig.water-equal} to reveal the effect of singular values' feedback on the transceiver design. Here, the singular values are assumed to be perfectly fed back to the Tx in PL-CSI and QL-CSI while completely unknown at the Tx for PLS-CSI and QLS-CSI. Hence, the optimal water-filling algorithm can be executed for PL-CSI and QL-CSI, and equal power allocation will be carried out for PLS-CSI and QLS-CSI. The transmit power is confined from $-20$ dBm to $-15$ dBm to avoid the high SNR regime, where equal power allocation is inherently asymptotically optimal. As anticipated, the water-filling algorithm achieves a higher ergodic SE compared with equal power allocation. However, considering that this SE gain is at the cost of accurate feedback of singular values, the tradeoff between performance gain and feedback overhead should be evaluated. Starting from the QLS-CSI with $B-B_{\min}=2$, two methods are used to implement this evaluation when more spare feedback bits are available, (a) feeding back singular values for water-filling power allocation at the Tx and (b) increasing the quantization accuracy of \textcolor[rgb]{0,0,0}{directional parameters} for channel customization at the RISs. When the transmit power equals $-19$ dBm, increasing the extra bits $B-B_{\min}$ to $4$ can achieve the same performance as the QL-CSI with $B-B_{\min}=2$, which requires accurate feedback for singular values. In other words, two more bits for \textcolor[rgb]{0,0,0}{directional parameters}' quantization in the QLS-CSI with $B-B_{\min}=2$ can yield a SE gain equivalent to infinite feedback for singular values. The benefit of \textcolor[rgb]{0,0,0}{directional parameters}' quantization is more pronounced with the transmit power. As discussed in Section \ref{sec:4.1}, the singular values of the customized channel are in the same order of magnitude by configuring the RISs' size and path pruning. Therefore, the superiority of the water-filling algorithm is compressed, and the equal power allocation can be asymptotically optimal. In this case, feedback overhead for singular values can be mitigated, which is especially critical when the Rx has limited feedback ability.

\begin{figure}[!t]
\centering
\includegraphics[width=0.5\textwidth]{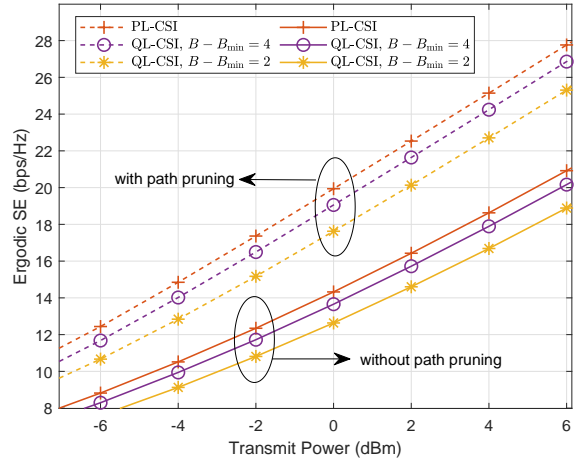}
\caption{Comparison of the ergodic SE with and without path pruning.}
\label{Fig.pruning} 
\end{figure}

Figure \ref{Fig.pruning} demonstrates the ergodic SE of the proposed channel customization-based SVD transceiver when orthogonal paths are selected with and without path pruning. The singular values of the composite channel determine the ergodic SE because it is customized in the form of SVD by the selected orthogonal paths. Thus, the paths with larger cascaded gain are selected to increase the singular values due to path pruning that omits weak paths, further improving the ergodic SE.\par

\begin{figure}[!t]
\centering
\includegraphics[width=0.5\textwidth]{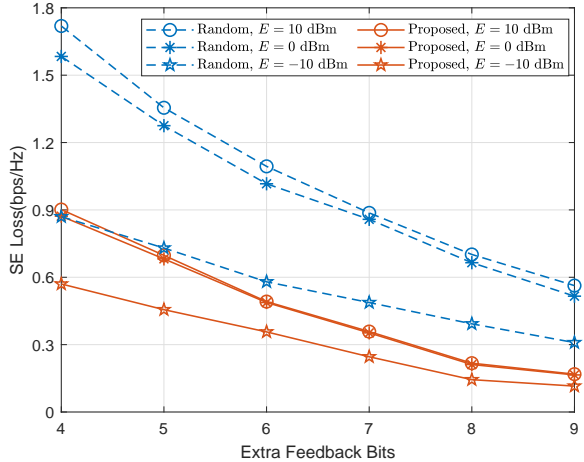}
\caption{Comparison of the SE loss under different bit allocation schemes.}
\label{Fig.RLvB} 
\end{figure}
The effects of extra feedback bits and corresponding bit allocation schemes on the SE loss are illustrated in Fig. \ref{Fig.RLvB}. With the guarantee of minimum feedback bits, random bit allocation is used as the baseline. The proposed bit partitioning algorithm can effectively reduce the SE loss when extra bits are limited. The narrower gap of SE loss indicates that the proposed bit partitioning strategy is robust against an increasing transmit power.\par

\begin{figure}[!t]
\centering
\includegraphics[width=0.5\textwidth]{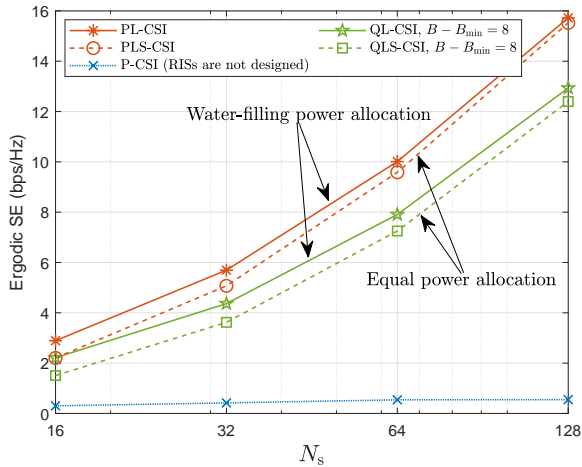}
\caption{Ergodic SE versus the number of RIS elements.}
\label{Fig.RvNs} 
\end{figure}
The numerical results in Figs. \ref{Fig.path-energy}--\ref{Fig.RLvB} are obtained with the RIS size constraint in \eqref{eq:N_s,k}, thereby ensuring that the singular values are of the same order of magnitude. To show the effect of the RIS size on the ergodic SE, we relax this limitation by setting $N_{{\rm S},{k}}=N_{\rm S}k^2$ with $N_{\rm S}$ increasing from $16$ to $128$. Figure \ref{Fig.RvNs} shows the ergodic SE versus $N_{\rm S}$ when the transmit power is $E=-10$ dBm. As shown in Fig. \ref{Fig.water-equal}, the ergodic SEs of PL-CSI and PLS-CSI are approximately the same at $E=-15$ dBm. However, they have approximately $1$ bps/Hz gap when $N_{{\rm S},{k}}=16k^2$ ($N_{\rm S}=16$) even at $E=-10$ dBm, as presented in Fig. \ref{Fig.RvNs}. From another point of view, this difference verifies that the RIS size setting in \eqref{eq:N_s,k} helps to create a well-conditioned channel with homogeneous singular values. Therefore, the simple equal power allocation can replace the water-filling algorithm in the low SNR regime. Figure \ref{Fig.RvNs} also shows that the ergodic SE increases with $N_{\rm S}$, and the performance gap between water-filling and equal power allocation is squeezed. This condition is because the received SNR can increase with $N_{\rm S}$ when the transmit power remains constant. However, the SE gap caused by the limited quantization is expanded with a larger $N_{\rm S}$. The underlying reason can be found in \eqref{eq:rate-loss-close} because the SE loss is an increasing function of ${\bar C}_k$ that grows with $N_{{\rm S},k}$. The significant performance gain over the P-CSI when the RISs are not designed emphasizes the substantial potential of channel customization via RIS configuration.


\section{Conclusion}\label{sec:6}
In this study, limited channel feedback was investigated for RIS-assisted FDD systems, where the Tx and Rx establish reliable transmission channels through multiple RISs. Given that large-scale RISs are deployed in the propagation environment between the Tx and Rx, directly feeding back the low-dimensional composite channel matrix or the high-dimensional individual subchannel matrices is impractical. Channel parameters' feedback is a feasible solution but entails huge overhead due to the rich-scattering environment. To reduce the feedback overhead, we proposed a channel customization scheme that reshapes the composite channel to be sparse. On the basis of channel customization, the optimal SVD transceiver can be obtained with limited CSI. Considering the limited feedback ability of the Rx, the effect of the parameter quantization on the SE loss was assessed. \textcolor[rgb]{0,0,0}{To reduce the SE loss, a bit partitioning strategy that allocates bits to quantize directional parameters was developed utilizing the derived closed-form expression of the SE loss.} The numerical results illustrated the potential of the channel customization in cutting down the feedback overhead and verified the efficiency of the proposed bit partitioning for reducing the SE loss.\par
\textcolor[rgb]{0,0,0}{This work is an early attempt on the limited channel feedback for multi-RIS-assisted systems. We moved away from existing works, which aim to feed back the complete channel parameters of every paths. Instead, we tailed the rich scattering environment into a sparse form and fed back only the dominant paths, which significantly reduces the feedback overhead. Moreover, utilizing the limited CSI to customize the composite channel in terms of SVD, we simplified the SVD transceiver by removing the matrix decomposition. We note that in the context of customizing channel characteristics, more research directions can be identified. For example, by using the statistical CSI that is spatially reciprocal for the uplink and downlink to successively enhancing paths, the downlink channel estimation problem can be simplified with more distinguishable paths. Moreover, by reshaping a block diagonal channel, the user interference can be mitigated without complicated transceiver design in RIS-assisted multi-user systems.}
\appendices

\section{}\label{App:A}

The constraint for the RIS size in \eqref{eq:N_s,k} enables the singular values to be in the same order of magnitude. Thus, we have the following approximation:
\begin{equation}\label{eq:A-1}
  \sum\limits_{i = 1}^{{N_{\rm{R}}}} {\frac{1}{{{N_{\rm{R}}}{{\left| {\xi _i^\star } \right|}^2}}}}  \approx \sum\limits_{i = 1}^{{N_{\rm{R}}}} {\frac{1}{{{N_{\rm{R}}}{{\left| {\xi _k^\star } \right|}^2}}}}  = \frac{1}{{{{\left| {\xi _k^\star } \right|}^2}}}.
\end{equation}
Then, the expectation of $C_k$ can be expressed as
\begin{equation}\label{eq:A-2}
\begin{aligned}
  {\mathbb E}\left\{{C_k}\right\} & = {\mathbb E}\left\{{\left| {\xi _k^\star } \right|^2}\left( {\frac{E}{{{N_{\rm{R}}}{\sigma ^2}}} + \sum\nolimits_{i = 1}^{{N_{\rm{R}}}} {\frac{1}{{{N_{\rm{R}}}{{\left| {\xi _i^\star } \right|}^2}}}}  - \frac{1}{{{{\left| {\xi _k^\star } \right|}^2}}}} \right)\right\}\\
   &\approx {\frac{E}{{{N_{\rm{R}}}{\sigma ^2}}} }{\mathbb E}\left\{{\left| {\xi _k^\star } \right|^2} \right\}.
\end{aligned}
\end{equation}
Expanding ${\xi _k^\star }$ and using the property of complex Gaussian variables, we have
\begin{equation}\label{eq:A-3}
\begin{aligned}
  {\mathbb E}\left\{{C_k}\right\} & \approx {\frac{E}{{{N_{\rm{R}}}{\sigma ^2}}} } {\mathbb E}\left\{{\left| {\xi _k^\star } \right|^2} \right\}
  = {\frac{E}{{{N_{\rm{R}}}{\sigma ^2}}} }  {\mathbb E}\left\{{\left| {{\rho _k}{\alpha _{{k,{\rm{R}}}}}{\alpha _{{\rm{T}},k,0}}} \right|^2} \right\} \\
  & ={\frac{E}{{{N_{\rm{R}}}{\sigma ^2}}} }   {\mathbb E}\left\{\frac{{{N_{\rm{T}}}{N_{\rm{R}}}N_{{\rm{S}},k}^2\rho _k^2{\kappa _{{\rm{T}},k}}}}{{{L_{{k,{\rm{R}}}}}\left( {{\kappa _{{\rm{T}},k}} + 1} \right)}}{\left| {{\beta _{{k,{\rm{R}}}}}} \right|^2}\right\}\\
   &={\frac{E}{{{N_{\rm{R}}}{\sigma ^2}}} }   \frac{{{N_{\rm{T}}}{N_{\rm{R}}}N_{{\rm{S}},k}^2\rho _k^2{\kappa _{{\rm{T}},k}}}}{{{L_{{k,{\rm{R}}}}}\left( {{\kappa _{{\rm{T}},k}} + 1} \right)}}.
\end{aligned}
\end{equation}
With the expressions of $N_{{\rm S},{k}}$ and $\rho_k$, we obtain
\begin{equation}\label{eq:A-4}
  {\mathbb E}\left\{{C_k}\right\}  \approx {\frac{E}{{{N_{\rm{R}}}{\sigma ^2}}} } {\left( {\frac{{d_{2,k}^{\rm{r}}\xi _0^\star }}{{{d_{2,k}}}}} \right)^2}.
\end{equation}



\begin{small}

\end{small}

\end{document}